\documentclass[preprint,12pt]{elsarticle}
\usepackage[table,svgnames]{xcolor}
\definecolor{Mahogany}{RGB}{192,64,0}
\definecolor{White}{RGB}{255,255,255}
\definecolor{Bittersweet}{RGB}{254,111,94}

\usepackage{amssymb}
\usepackage{amsmath}

\usepackage{booktabs}   
\usepackage{multirow}   
\usepackage{url}
\usepackage[colorlinks=true,linkcolor=blue,citecolor=blue,urlcolor=blue]{hyperref}

\usepackage{algorithm}
\usepackage{algorithmic}

\begin{document}

\begin{frontmatter}



\title{Discrete Elastic Ribbons: A Unified Discrete Differential Geometry Framework for One-Dimensional Energy Models}


\author{Shivam K Panda, M Khalid Jawed} 

\affiliation{organization={University of California, Los Angeles},
            addressline={420 Westwood Plaza}, 
            city={Los Angeles},
            postcode={90095}, 
            state={CA},
            country={United States of America}}

\begin{abstract}
Elastic ribbons, slender structures whose length ($L$), width ($W$), and thickness ($b$) satisfy $L \gg W \gg b$, exhibit mechanical behaviors intermediate between one-dimensional rods ($L \gg W, b$) and two-dimensional plates ($L, W \gg b$). In quadratic Kirchhoff-type rod-based frameworks, such as Discrete Elastic Rods (DER), the governing equilibrium equations are independent of width, and therefore these models cannot capture width-dependent mechanical effects. Reduced centerline-based ribbon models attempt to capture width dependence via coupled bending–twisting energies. However, their relative accuracy remain unclear due to the absence of a unified simulation framework. In this work, we formulate a framework grounded in discrete differential geometry where the energy is expressed as functions of coupled bending–twisting strain measures along the centerline, rather than a linear sum of quadratic bending and twisting energies in DER. We derive analytical gradients and Hessians of the energy that enable implicit time integration. Within this unified setting, we compare five ribbon models: Kirchhoff, Sadowsky, Wunderlich, Sano, and Audoly. As a benchmark, a straight ribbon is longitudinally constrained into a pre-buckled arch and subjected to transverse displacement, inducing a supercritical pitchfork bifurcation. Predicted bifurcation thresholds are compared against shell-based finite element simulations, with the Sano model providing the closest agreement in capturing width-dependent shifts. Our high-performance JAX-based implementation achieves $\mathcal{O}(N)$ per-iteration cost and also confirms that Sano model introduces negligible per-iteration overhead relative to standard DER.
\end{abstract}


\end{frontmatter}




\section{Introduction}
\label{sec:intro}

Thin elastic structures appear across a broad range of natural and engineered systems, from plant tendrils and DNA supercoiling to biomedical and molecular devices \cite{marras2015programmable}, stretchable electronics \cite{yang2017elasticity, stanley2024stretchable}, deployable aerospace structures \cite{seffen1999deployment, lappas2011cubesail}, and soft robotic actuators \cite{liu2025electro, chen2018harnessing}. Among these, elastic ribbons are distinguished by a dimensional hierarchy in which the length $L$ greatly exceeds the width $W$, which itself greatly exceeds the thickness $b$ ($L \gg W \gg b$), placing them at the intersection of one-dimensional rod theory and two-dimensional plate mechanics. This intermediate character endows ribbons with rich phenomenology including snap-through instabilities, twist-induced shape transitions, dynamic mode skipping and selection \cite{huang2024exploiting}, and complex multistable configurations that have attracted sustained interest from both the mechanics and applied mathematics communities \cite{dias2015wunderlich,starostin2007shape,sano2019twist}.

The mechanics of ribbons can be approached from two distinct perspectives. From the plate viewpoint, ribbons are narrow elastic surfaces.
While plate models faithfully resolve two-dimensional deformation fields, discretizing a ribbon with shell or plate elements requires a fine mesh along the width direction even though the dominant kinematics are one-dimensional, making such simulations computationally expensive. From the rod viewpoint, ribbons are slender structures whose behavior can be captured by tracking only the centerline and an attached material frame. These models are orders of magnitude cheaper computationally, however accurately capturing width dependence and coupled bending-twisting mechanics along the centerline is a key challenge. The practical appeal of one-dimensional models, namely their computational efficiency and analytical tractability, has motivated considerable effort towards developing reduced ribbon theories that encode the essential physics of the two-dimensional surface into centerline quantities.

The earliest such reduction was proposed by Sadowsky~\cite{sadowsky1930elementarer}, who derived an energy functional for narrow inextensible ribbons by enforcing developability, i.e., vanishing Gaussian curvature. Under this constraint, the ribbon may bend but cannot have transverse surface curvature across its width; locally, the surface can curve in only one principal direction. Equivalently, a developable ribbon is a surface ruled by straight generators whose direction may vary along the centerline; the evolution of this ruling direction induces the coupling between bending and twist. In a surface-adapted frame, this eliminates transverse bending and leaves bending primarily along the ribbon centerline. However, the direction of bending may rotate along the strip, and this rotation is geometrically coupled to twisting of the centerline, so that twist and bending are no longer independent. The Sadowsky functional captures this coupling through a penalty term that diverges when twisting remains finite while bending vanishes, thereby enforcing developability within a reduced one-dimensional energy. Wunderlich~\cite{wunderlich1962abwickelbares} subsequently generalized this approach to account for finite ribbon width, deriving an energy that depends on the spatial derivatives of the twist-to-curvature ratio. Although this functional provides higher geometric fidelity, it comes at the cost of logarithmic singularities at inflection points, where the centerline curvature vanishes while the twisting of the centerline remains nonzero, making numerical implementation challenging~\cite{starostin2015equilibrium,moore2019computation}.

In contrast, a simpler alternative is to model the ribbon as a Kirchhoff rod with strongly anisotropic bending stiffness, reflecting the much greater resistance to bending about the wide axis than about the thin axis of its rectangular cross-section. While computationally convenient, this approach fails to capture the characteristic stiffening behavior of ribbons at large deformations where the coupling between bending and twisting becomes significant. More recently, Sano and Wada~\cite{sano2019twist} (hereafter referred to as the Sano model) proposed a regularized Sadowsky model that introduces a width-dependent parameter to prevent divergence at inflection points, enabling stable numerical simulation while preserving the essential physics in high-curvature regions. 
Audoly and Neukirch~\cite{audoly2021one} (hereafter referred to as the Audoly model) derived an asymptotic one-dimensional ribbon model from plate theory. Their formulation incorporates finite-width effects and introduces a transition function that smoothly connects Kirchhoff rod-like behavior with the developable Sadowsky limit, accounting for mid-surface stretching and anticlastic (Poisson) effects that become relevant at intermediate curvatures.

Despite this variety of ribbon models, their relative performance remains poorly characterized due to the absence of a unified numerical framework capable of simulating all formulations under identical conditions. Previous computational studies have typically focused on individual models: continuation methods have been applied to the Sadowsky functional \cite{starostin2007shape,starostin2015equilibrium}, elliptic regularization has enabled Wunderlich simulations \cite{moore2019computation}, curvature-based elements have achieved high accuracy for inextensible ribbons \cite{charrondiere2020numerical,charrondiere2024merci}, purely numerical computational frameworks have been developed to derive effective one-dimensional constitutive laws from cross-sectional analysis of strip geometry \cite{KUMAR2024116553}, and a ruled-surface regularization with an auxiliary field has been recently proposed to resolve Sadowsky singularities at inflection points \cite{vitral2025ruled}. Huang et al.\ \cite{huang2023bifurcations} reformulated Discrete Elastic Rods (DER) as a static solver equipped with eigenvalue analysis of the tangential stiffness matrix, enabling detection of bifurcation points and branch switching in beams, strips, and gridshells, but it is limited to only the Kirchhoff constitutive law without extending to any ribbon-specific energy models for strips. All of these approaches employ different numerical strategies, boundary condition implementations, and convergence criteria, precluding direct comparison of the underlying physical models. Furthermore, the numerically stiff character of ribbon equations, particularly the singular behavior of developable models at inflection points, has limited most previous work to quasi-static loading scenarios or required manual intervention to traverse critical configurations. The closest precursor to our work is Huang et al.\ \cite{huang2022discrete}, who developed a discrete differential geometry (DDG)-based framework that unifies three one-dimensional energy models, namely Kirchhoff, Sadowsky, and a simplified Audoly-Neukirch formulation, as limiting cases of a single extensible ribbon constitutive law discretized within the DER algorithm. However, that study was validated mostly at narrow widths, leaving the performance of these energy models at wider ribbons uncharacterized. Moreover, the framework is tailored to that specific constitutive form, so incorporating other ribbon energy models, such as the full Wunderlich functional with its logarithmic terms, or the complete Audoly-Neukirch formulation with its transition function, would require substantial model-specific reformulation rather than straightforward substitution.

The need for systematic model comparison is highlighted by the work of Huang et al.\ \cite{huang2020shear}, who studied the shear-induced supercritical pitchfork bifurcation of pre-buckled bands using both one-dimensional rod and two-dimensional plate simulations validated against experiments. This benchmark problem reveals that the anisotropic Kirchhoff rod accurately predicts bifurcation behavior only for narrow strips ($W/L \lesssim 1/20$), while the plate model maintains accuracy even when $W/L \sim 1/2$. Discrete Sadowsky simulations were shown to fail at capturing the bifurcation due to the energy barrier at inflection points. This raises a critical open question: which one-dimensional ribbon model, if any, can bridge the gap between rod and plate predictions across the full range of ribbon geometries?

In this paper, we address this question by developing the first numerical framework capable of simulating all prominent one-dimensional ribbon energy models: (i) Kirchhoff, (ii) Sadowsky, (iii) Wunderlich, (iv) Sano, and (v) Audoly, within a unified DDG formulation. Following the successful DER methodology \cite{bergou2008discrete,bergou2010discrete,jawed2018primer}, we discretize the ribbon centerline into nodal positions and edge-based twist angles, defining curvature and twist at internal vertices through geometric operations that naturally preserve frame invariance. The key innovation of our approach is a generalized energy formulation that expresses the elastic potential as a function of the strain vector $\boldsymbol{\epsilon} = [\varepsilon, \kappa^{(1)}, \kappa^{(2)}, \tau]^T$, enabling systematic derivation of gradients and Hessians regardless of the specific constitutive model. Beyond the unified framework, we investigate the effect of ribbon thickness on the bifurcation response, a parameter not previously studied within the one-dimensional modeling context, by examining two length-to-thickness ratios ($L/b = 100$ and $L/b = 1000$) that enter each energy model through distinct mechanisms. We further introduce a homotopy-based width transition strategy that enables the discovery of bifurcation branches at large widths where direct quasi-static loading fails. To assess computational overhead, we also develop a high-performance JAX reimplementation and benchmark per-step costs across energy models, demonstrating that the coupled ribbon models, especially Sano, introduce negligible overhead relative to the standard Kirchhoff-based DER.

Our simulation framework\footnote{Source code is available at \url{https://github.com/StructuresComp/discrete-elastic-ribbon}.} leverages the open-source DER implementation DisMech~\cite{lahoti2025py} to support the generalized energy formulation of Section~\ref{sec:generalized_energy} within a robust implicit time-stepping scheme. Although the problems considered are primarily quasi-static, we employ a fully implicit Euler integration method, which introduces inertial terms that regularize the equilibrium solve and improve numerical stability near instabilities. Newton-Raphson (NR) iteration with exact Jacobians achieves quadratic convergence, while adaptive time-stepping automatically refines the temporal resolution near bifurcation points and snap-through events. To handle the numerical stiffness inherent to developable ribbon models, particularly the near-singular Hessians encountered when the transverse curvature approaches zero, we incorporate Tikhonov-Miller regularization that ensures positive definiteness while introducing negligible perturbation to the converged solution. This combination of techniques enables robust simulation of all ribbon models, including the Wunderlich functional with its challenging logarithmic terms, on the same boundary value problem.

We adopt the DER framework, which belongs to the family of geometrically exact discrete formulations for slender structures~\cite{bergou2008discrete, bergou2010discrete, jawed2018primer}. In the sense introduced by
Simo~\cite{simo1985finite, simo1989stress, simo1990stress1, simo1990stress2}, a formulation is geometrically exact when its governing equations are valid for arbitrarily large displacements and rotations without geometric approximation beyond the kinematic assumption itself. DER achieves this through discrete differential geometry: the curvature binormal is computed from exact cross products of edge vectors, and the material frame evolves via parallel transport along the discrete centerline, ensuring that the resulting strain measures are frame-invariant by construction for arbitrary deformation magnitudes. Alternative discrete beam models, such as the Piola--Hencky elastic chain
framework of Turco~\cite{turco2018discrete}, define energy through angular spring potentials between rigid bars rather than through strain measures derived from the differential geometry of the discrete curve, and therefore do not share this geometric exactness.

A second consideration specific to ribbon mechanics is constitutive compatibility. All five ribbon energy models considered in this work are formulated in terms of stretching, bending curvatures, and twist,
$\boldsymbol{\epsilon} = [\varepsilon, \kappa^{(1)}, \kappa^{(2)}, \tau]^T$, without transverse shear. DER computes precisely these strain measures from its discrete centerline and material frame, then
passes them as a self-contained vector to the energy function $\mathcal{E}(\boldsymbol{\epsilon})$. This clean interface is what enables the generalized framework of Section~\ref{sec:generalized_energy}, where swapping ribbon energy models requires only changing $\mathcal{E}(\boldsymbol{\epsilon})$ with no modification to the geometric derivative framework. Cosserat rod~\cite{gazzola2018forward}, by contrast, include transverse shear strains that are not utilized by any of the ribbon energy models. DER also offers computational advantages for the implicit time integration required by the stiff ribbon energy landscapes studied in our work. Its time-parallel transport updates the material frame in $\mathcal{O}(N)$ without a global solve, and its local element stencil yields a fixed-bandwidth banded Jacobian that enables $\mathcal{O}(N)$ factorization per Newton iteration. A recent comparison~\cite{lahoti2025py} confirmed that these properties make DER more computationally efficient than a Cosserat rod-based framework for equivalent beam problems.

\begin{figure}[ht]
    \centering
    \includegraphics[width=1.0\textwidth]{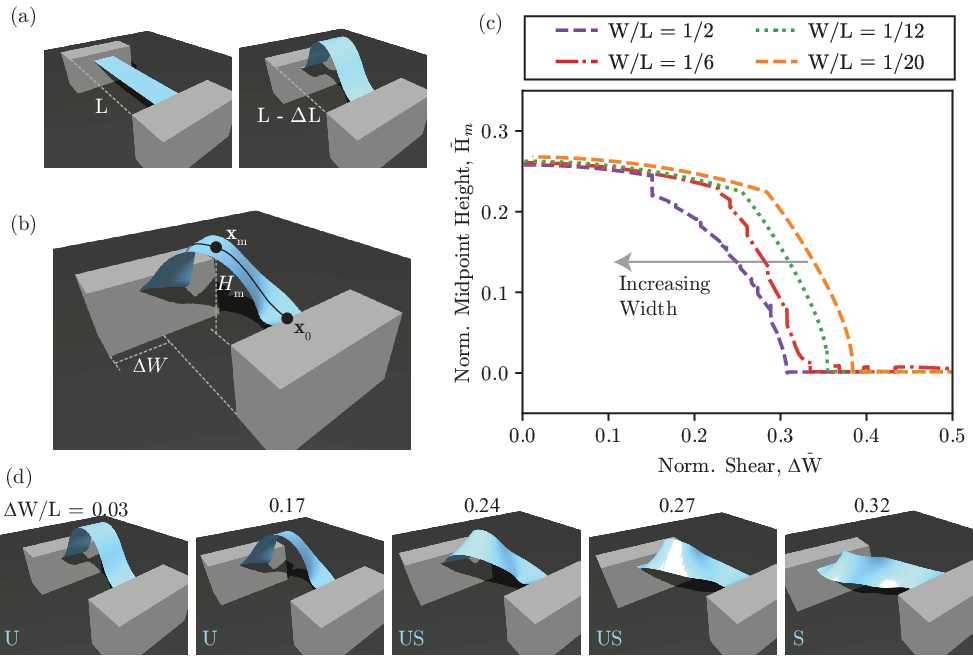}
    \caption{Overview of the shear-induced pitchfork bifurcation with results for Sano energy model. (a) Longitudinal compression $\Delta L$ inducing a buckled equilibrium. (b) Transverse shear $\Delta W$ on clamped end inducing supercritical pitchfork bifurcation. (c) Sano energy model's performance assessed using normalized midpoint height $\bar{H}_m = |H_{m}|/L$ vs normalized transverse shear $\Delta \bar{W} = \Delta W/L$ curve in capturing the leftward shift of bifurcation transition points on increasing width. (d) Snapshots from renderings of $W/L=1/6$ width as transverse shear is varied by quasi-static and we observe $U$, $US$ \& $S$ shapes.}
    \label{fig:overview}
\end{figure}

We validate our framework against the shear-induced bifurcation benchmark of Huang et al.\ \cite{huang2020shear}, which provides experimental data and plate simulation results for comparison. The boundary conditions, clamped ends with prescribed longitudinal compression (Figure~\ref{fig:overview}a), followed by lateral shear (Figure~\ref{fig:overview}b), induce a sequence of configuration changes from symmetric $U$-shaped to asymmetric $US$ and $S$-shaped patterns (Figure~\ref{fig:overview}d), with supercritical pitchfork bifurcations marking the transitions. Our systematic comparison reveals that the Sano model achieves the highest accuracy among the five one-dimensional formulations, consistently capturing the width-dependent leftward shift of bifurcation thresholds (Figure~\ref{fig:overview}c) while maintaining the lowest absolute errors relative to finite element analysis (FEA) across $W/L \in \{1/20, 1/12, 1/6\}$ (refer to Figure~\ref{fig:all-energies-hmid-b1e-3}(a-c) and Supplementary Videos S1-S3). Nevertheless, the predicted leftward shifts fall short of the FEA reference, with the discrepancy growing as the ribbon widens and exposing fundamental limitations of such analytical energy models in representing genuinely two-dimensional deformation mechanisms. The Audoly model shows comparable accuracy at specific width-thickness combinations but lacks consistent performance across the full parameter space, while the strict developable models, Sadowsky and Wunderlich, cannot spontaneously traverse the pitchfork bifurcation under any conditions tested. At large widths ($W/L \geq 1/3$), where all one-dimensional models fail to bifurcate under direct loading, our homotopy-based width transition successfully uncovers hidden bifurcation branches within the Sano energy landscape, demonstrating that the model supports the correct equilibrium structure even when the solver cannot reach it through standard quasi-static loading (refer to Figure~\ref{fig:homotopy-width-sano} and Supplementary Videos S5-S6). These results establish the Sano model as the current best-performing one-dimensional baseline, superseding the Kirchhoff rod for ribbons beyond the narrow-strip limit, and motivate the development of improved ribbon energy formulations that can more faithfully encode width-dependent mechanics. To test whether these conclusions generalize beyond a single loading scenario, we further validate the framework on two additional benchmarks: boundary twist (refer to Figure~\ref{fig:twist-bc} and Supplementary Videos S10-S11) and combined shear-plus-twist loading (refer to Figure~\ref{fig:shear-plus-twist-bc} and Supplementary Videos S12-S13) of pre-buckled bands. The Sano model's superior performance persists across all three loading conditions.

The remainder of this paper is organized as follows. Section~\ref{sec:ribbonModels} summarizes the five centerline-based ribbon models considered in this paper. Section~\ref{sec:kinematics} presents the kinematic representation of the ribbon within a DDG framework. Section~\ref{sec:generalized_energy} develops a generalized discrete energy formulation and derives explicit gradient and Hessian expressions applicable to nonlinear energy models with coupling between strain measures such as bending and twisting. Section~\ref{sec:models} describes the discrete implementations of the ribbon energy models and presents their DDG counterparts. Section~\ref{sec:results} presents systematic comparisons on the shear-induced bifurcation problem. Section~\ref{sec:cost} analyzes the computational cost of the framework relative to standard DER. Section~\ref{sec:relevant_benchmarks} extends the validation to twist and combined shear-plus-twist loading conditions. Conclusions finally follow in Section~\ref{sec:conclusion}.

\section{Ribbon Energy Models}
\label{sec:ribbonModels}

\begin{figure}[b!]
    \centering
    \includegraphics[width=1\textwidth]{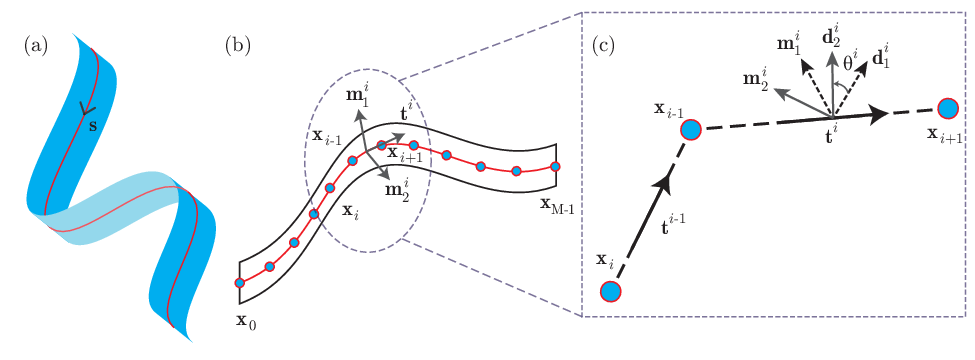}
    \caption{(a) Ribbon parametrized by arc length $s$. (b) Discrete representation of the ribbon. (c) Notation used in the discrete model, including nodal coordinates, material frame, and reference frame.}
    \label{fig:kinematics}
\end{figure}

Referring to Figure~\ref{fig:kinematics}a, the centerline of the ribbon is parameterized by the arc-length variable $s \in [0,L]$. The ribbon possesses a rectangular cross-section of width $W$ and thickness $b$ with $L \gg W \gg b$. The material is assumed to be homogeneous and linearly elastic, with Young’s modulus $E$ and shear modulus $G$. At each point along the centerline, we attach a material frame $\{\mathbf{m}_1(s), \mathbf{m}_2(s), \mathbf t (s)\}$, where $\mathbf t$ is tangent to the centerline and $\mathbf{m}_1$ is chosen normal to the ribbon surface. The remaining director $\mathbf{m}_2 = \mathbf t \times \mathbf{m}_1$ lies within the surface and spans the ribbon width. The evolution of this material frame along $s$ defines three strain measures: the bending curvatures $\kappa^{(1)}(s)$ and $\kappa^{(2)}(s)$ corresponding to rotations about $\mathbf{m}_2$ and $\mathbf{m}_1$, and the twist $\tau(s)$ corresponding to rotation about the centerline tangent $\mathbf t$.

The sectional properties of the rectangular cross-section are
\begin{equation}
A = Wb, 
\qquad 
I_1 = \frac{bW^3}{12}, 
\qquad 
I_2 = \frac{Wb^3}{12}, 
\qquad 
J \approx \frac{Wb^3}{3},
\end{equation}
where $A$ is the cross-sectional area. The second moments of area $I_1$ and $I_2$ correspond to bending about the $\mathbf{m}_2$ and $\mathbf{m}_1$ directions, respectively. The torsional constant $J$ is given by the Saint-Venant approximation for thin rectangular sections with $W \gg b$. For thin ribbons, the inequality $I_1 \gg I_2$ represents the strong anisotropy of the cross-section: bending associated with $I_2$ (about the thin direction) is significantly less stiff than bending associated with $I_1$ (about the wide direction).

In this section, we summarize the classical one-dimensional energy formulations for elastic ribbons in their continuous form. The corresponding energies $E^{(\cdot)}$ for each model are listed in Table~\ref{tab:ribbon_energies}, where the superscript denotes the specific ribbon formulation. These continuous expressions are the foundations for the discrete implementations introduced later in Section~\ref{sec:models}. With the exception of the classical Kirchhoff rod, all models incorporate explicit coupling between bending and twisting strains.

\begin{table}[htbp]
\centering
\small
\renewcommand{\arraystretch}{2.8}
\begin{tabular}{@{} p{0.12\linewidth} p{0.82\linewidth} @{}}
\hline
\textbf{Model} & \textbf{Energy Functional} $\mathcal{E}$ \\[0.5ex]
\hline
Kirchhoff & 
$\displaystyle E^\textrm{Kirc} = \int_0^L \!\left[ EI_1 \left(\kappa^{(1)}\right)^2 + EI_2\left(\kappa^{(2)}\right)^2 + GJ\tau^2 \right] ds$ \\[2ex]

Wunderlich & 
$\displaystyle E^\textrm{Wund} = \int_0^L \!EI_2 \left[\kappa^{(2)}(1+\eta^2)\right]^2 \frac{1}{W\eta'} \log \!\left( \frac{1 + \frac{W\eta'}{2}}{1 - \frac{W\eta'}{2}} \right) ds$
\newline
where $\eta = \tau/\kappa^{(2)}$ and $(\cdot)'$ denotes differentiation with respect to $s$ \\[2ex]

Sadowsky & 
$\displaystyle E^\textrm{Sado} = \int_0^L \!\left[ EI_2(\kappa^{(2)})^2 + GJ\tau^2 + EI_2 \,\frac{\tau^4}{\left(\kappa^{(2)}\right)^2} \right] ds$ \\[2ex]

Sano & 
$\displaystyle E^\textrm{Sano} = \int_0^L \!\left[ EI_2(\kappa^{(2)})^2 + GJ\tau^2 + EI_2 \,\frac{\tau^4}{1/\zeta^2 + \left(\kappa^{(2)}\right)^2} \right] ds$ \newline
where $\zeta^2 = \dfrac{(1-\nu)W^4}{60b^2}$ \\[2ex]

Audoly & 
$\displaystyle 
\begin{aligned}
E^\textrm{Aud} = \int_0^L \!\Bigg[
& EI_2(\kappa^{(2)})^2 + GJ\tau^2 \\
& +\, 6 EI_2 \left( \frac{W^2}{b}\right)^2 
\left( \nu (\kappa^{(2)})^2 + \tau^2 \right)^2 
\varphi\!\left( \frac{\kappa^{(2)}}{\kappa^*} \right)
\Bigg] ds
\end{aligned}
$ \newline
where $\kappa^* = \frac{1}{\sqrt{12(1-\nu^2)}} \frac{b}{W^2}$ \\[0.5ex]
\hline
\end{tabular}
\caption{Summary of continuous ribbon energy models. $W$ and $b$ denote width and thickness, respectively. Here we are comparing inextensible terms that define the bending-twisting coupling.}
\label{tab:ribbon_energies}
\end{table}

\textbf{Kirchhoff Anisotropic Rod Model:} The classical Kirchhoff model serves as the baseline, treating the ribbon as an anisotropic rod. While it accounts for the disparate bending stiffnesses of the cross-section ($EI_1 \gg EI_2$), it does not explicitly capture the coupling between bending and twisting induced by the ribbon's width. Consequently, it fails to enforce the geometric constraint of developability, making it accurate only for narrow strips \cite{yu2019bifurcations} or small deformations.

\textbf{Wunderlich Ribbon Model:} To address the finite width $W$, Wunderlich treated the ribbon as a developable ruled surface \cite{wunderlich1962abwickelbares}. This formulation integrates the elastic energy density across the width, resulting in a highly nonlinear functional that depends on the derivatives of the curvature and torsion. Although it offers high geometric fidelity, the presence of logarithmic terms and higher-order derivatives ($\eta'$) makes it analytically complex and computationally expensive for numerical simulations \cite{moore2019computation}.

\textbf{Sadowsky Ribbon Model:} For long, narrow ribbons ($L \gg W$) undergoing large deformations, the Wunderlich model reduces to the Sadowsky limit \cite{sadowsky1930elementarer}. This model enforces developability through a singular penalty term: $EI_2 \tau^4 / (\kappa^{(2)})^2$, where the energy diverges to infinity if the ribbon twists ($\tau \neq 0$) while the minor-axis curvature ($\kappa^{(2)}$) vanishes. This singularity strictly forbids the formation of inflection points with non-zero twist, which, while geometrically rigorous for inextensible surfaces, poses significant stability challenges for numerical minimization.

\textbf{Sano Ribbon Model:} Sano \& Wada \cite{sano2019twist} proposed a regularization of the Sadowsky energy to mitigate its numerical instability. By introducing a ``soft'' constraint parameter $\zeta$, derived from the cross-sectional geometry, the model prevents divergence at inflection points ($\kappa^{(2)} \to 0$). This modification preserves the physics of developability in high-curvature regions while allowing the solver to traverse straight configurations smoothly.

\textbf{Audoly Ribbon Model:} Bridging the gap between the rod-like behavior of the Kirchhoff model and the surface-like behavior of the Sadowsky model, Audoly \& Neukirch \cite{audoly2021one} derived an asymptotic formulation from 2D plate theory. Their model accounts for mid-surface stretching and anticlastic curvature (Poisson effects), using a transition function $\varphi$ to interpolate between regimes. It recovers the Kirchhoff energy at small strains and the Sadowsky energy at large strains, offering a unified description of the ribbon's stiffening behavior. Furthermore, this provides a theoretical basis for the heuristic regularization proposed by Sano \& Wada, which is recovered in the limit of vanishing Poisson's ratio ($\nu=0$) when a rational approximation is used for the transition function.

\section{Kinematics}
\label{sec:kinematics}

This section presents the discrete kinematic representation of the ribbon based on the DDG formulation of DER~\cite{bergou2008discrete,jawed2018primer}. We define the strain measures and their dependence on the degrees of freedom; the constitutive energy model, including any coupling between strain components, is introduced later in Section~\ref{sec:generalized_energy}.

\subsection{Discrete Geometric Representation}
\label{sec:discreteGeometricRepresentation}

Referring to Figure~\ref{fig:kinematics}, the ribbon centerline (Figure~\ref{fig:kinematics}a) is approximated through a discrete representation (Figure~\ref{fig:kinematics}b) consisting of $M$ nodal points $\{\mathbf{x}_0, \mathbf{x}_1, \ldots, \mathbf{x}_{M-1}\}$. Adjacent nodes are connected via $(M-1)$ edge vectors defined as
\begin{equation}
\mathbf{e}^i = \mathbf{x}_{i+1} - \mathbf{x}_i, \quad i = 0, 1, \ldots, M-2.
\end{equation}

As shown in Figure~\ref{fig:kinematics}(c), the local coordinate system at each edge is characterized by two orthonormal frames. The \textit{reference frame} $\{\mathbf{d}_1^i, \mathbf{d}_2^i, \mathbf{t}^i\}$ consists of the unit tangent $\mathbf{t}^i = \mathbf{e}^i / \|\mathbf{e}^i\|$ and orthonormal directors $\mathbf{d}_1^i, \mathbf{d}_2^i \perp \mathbf{t}^i$. The \textit{material frame} $\{\mathbf{m}_1^i, \mathbf{m}_2^i, \mathbf{t}^i\}$ is obtained through rotation of the reference frame by twist angle $\theta^i$
\begin{equation}
\mathbf{m}_1^i = \cos(\theta^i)\mathbf{d}_1^i + \sin(\theta^i)\mathbf{d}_2^i, \qquad \mathbf{m}_2^i = \mathbf{t}^i \times \mathbf{m}_1^i.
\end{equation}
The material frame orientation aligns $\mathbf{m}_1^i$ with the original plate's surface normal direction.

The complete rod configuration is described by a $(4M-1)$-dimensional state vector
\begin{equation}
\mathbf{q} = \left[\mathbf{x}_0^T, \mathbf{x}_1^T, \ldots, \mathbf{x}_{M-1}^T, \theta^0, \theta^1, \ldots, \theta^{M-2} \right]^T,
\end{equation}
including both translational coordinates and rotational degrees of freedom.

\subsection{Discrete Strain Measures}

The discrete configuration induces strain measures defined on edges and nodes. 
Axial strains are associated with edges, while bending curvatures and twist are associated with {\em interior} nodes that have two adjacent edges (i.e., nodes $i = 1, \ldots, M-2$). These strain measures provide a frame-invariant geometric description of deformation and serve as inputs to the constitutive energy models introduced later.

The axial strain is defined on each edge $i = 0, \ldots, M-2$ as
\begin{equation}
\varepsilon^i = \frac{\|\mathbf{e}^i\|}{\|\overline{\mathbf{e}}^i\|} - 1,
\end{equation}
where $\|\overline{\mathbf{e}}^i\|$ denotes the undeformed edge length, yielding $(M-1)$ axial strain values.

Bending strains are defined at nodes $i = 1, \ldots, M-2$. The discrete curvature is computed via the curvature binormal vector
\begin{equation}
(\boldsymbol{\kappa}_b)_i = \frac{2\mathbf{e}^{i-1} \times \mathbf{e}^i}{\|\mathbf{e}^{i-1}\|\|\mathbf{e}^i\| + \mathbf{e}^{i-1} \cdot \mathbf{e}^i},
\end{equation}
Projection onto the material frame yields two curvature components
\begin{equation}
\kappa_i^{(1)} = \frac{1}{2}\left(\mathbf{m}_2^{i-1} + \mathbf{m}_2^i\right) \cdot (\boldsymbol{\kappa}_b)_i, 
\qquad 
\kappa_i^{(2)} = -\frac{1}{2}\left(\mathbf{m}_1^{i-1} + \mathbf{m}_1^i\right) \cdot (\boldsymbol{\kappa}_b)_i,
\end{equation}
yielding $(M-2)$ curvature values for each component.

The twist is also defined at nodes $i = 1, \ldots, M-2$,
\begin{equation}
\tau_i = \theta^i - \theta^{i-1} + m_{\text{ref}}^i,
\end{equation}
where $m_{\text{ref}}^i$ denotes the reference twist contribution, yielding $(M-2)$ twist values.

\section{Generalized Energy Formulation}
\label{sec:generalized_energy}

The classical Kirchhoff rod model assumes that the elastic energy is a ``linear" sum of terms, each being ``quadratic" in a single strain measure. While computationally convenient, this decomposition fails to capture the coupled bending-twisting behavior inherent to developable ribbons. Popular ribbon models including Sadowsky and Wunderlich, introduce nonlinear coupling between curvature and twist, necessitating a generalized framework.

\subsection{Energy as a Function of Strains}

The total elastic energy is assembled from element-level contributions
\begin{equation}
    E(\mathbf{q}) = \sum_{k=1}^{M-2} 
    \mathcal{E}_k\!\left(\{\boldsymbol{\epsilon}_j\}_{j \in \mathcal{N}(k)}\right),
    \label{eq:totalEnergy}
\end{equation}
where $\boldsymbol{\epsilon}_k = [\varepsilon_k, \kappa^{(1)}_k, \kappa^{(2)}_k, \tau_k]^T$ collects the four strain measures at element $k$, and $\mathcal{N}(k)$ denotes the set of element indices whose strains appear in $\mathcal{E}_k$. For \emph{local} models (Kirchhoff, Sadowsky, Sano, Audoly), $\mathcal{N}(k) = \{k\}$ and the element energy depends only on its own strains. For the \emph{non-local} Wunderlich model, $\mathcal{N}(k) = \{k-1, k, k+1\}$ because the spatial derivative $\eta'_k = (\tau/\kappa^{(2)})'$ (see Section~\ref{sec:discrete-wunderlich}) requires finite differences over neighboring elements.

This formulation embeds frame invariance: expressing energy through strain measures defined via projections onto the material frame ensures that $E$ is invariant under rigid-body transformations and the resulting forces are equivariant. The specific form of $\mathcal{E}_k$ for each model is given in Section~\ref{sec:models}.

\subsection{Gradient of Energy}

The gradient of the total energy with respect to the global DOFs at index set $\mathcal{I}_j$ is obtained by differentiating 
Equation~\eqref{eq:totalEnergy}
\begin{equation}
    \frac{\partial E}{\partial \mathbf{q}[\mathcal{I}_j]} = 
    \sum_{k \,:\, j \in \mathcal{N}(k)} 
    \mathbf{G}_j^T \,\frac{\partial \mathcal{E}_k}{\partial \boldsymbol{\epsilon}_j},
    \label{eq:globalGradient}
\end{equation}
where $\mathbf{G}_j = \partial\boldsymbol{\epsilon}_j/\partial\mathbf{q}[\mathcal{I}_j]$ 
is the $(4 \times 11)$ strain Jacobian at element $j$, whose analytical expressions are derived in ~\ref{app:strain_gradients}.

The internal force vector is assembled as $\mathbf{F}_\text{int} = -\nabla_\mathbf{q} E$ by scattering the contribution 
$-\mathbf{G}_j^T\,\partial\mathcal{E}_k/\partial\boldsymbol{\epsilon}_j$ into global slots $\mathcal{I}_j$ for every pair $(k,j)$ with $j \in \mathcal{N}(k)$.

For local models the sum collapses to a single term, recovering the standard element-wise chain rule
\begin{equation}
    \frac{\partial E}{\partial \mathbf{q}[\mathcal{I}_k]}
    \bigg|_{\mathcal{N}(k)=\{k\}} 
    = \mathbf{G}_k^T \,\frac{\partial \mathcal{E}_k}{\partial \boldsymbol{\epsilon}_k}.
\end{equation}

\subsection{Hessian of Energy}

Implicit time integration requires the Hessian $\mathbf{H}_E = \nabla^2_\mathbf{q} E$. The contribution of element $k$ to the Hessian block associated with its own 11 local DOFs is
\begin{equation}
    \mathbf{H}_E\big[\mathcal{I}_k, \mathcal{I}_k\big] = 
    \sum_{k=1}^{M-2}\left(
    \underbrace{\mathbf{G}_k^T 
    \!\left(\nabla^2_{\boldsymbol{\epsilon}_k} \mathcal{E}_k\right)\!
    \mathbf{G}_k}_{\text{constitutive stiffness}} \;+\; 
    \underbrace{\sum_{l=1}^{4} \frac{\partial \mathcal{E}_k}{\partial \epsilon_{k,l}} 
    \left[\mathbf{H}_{\epsilon,k}\right]_l}_{\text{geometric stiffness}}
    \right),
    \label{eq:localHessian}
\end{equation}
where $\mathbf{H}_{\epsilon,k} = \partial^2\boldsymbol{\epsilon}_k/ \partial\mathbf{q}[\mathcal{I}_k]^2$ is the $(4\times 11\times 11)$ geometric Hessian of the strain map at element $k$, derived in ~\ref{app:hessian_derivation}.

The constitutive stiffness term encodes how energy curves in strain space. For the Kirchhoff model, $\nabla^2_{\boldsymbol{\epsilon}_k}\mathcal{E}_k$ is diagonal and the standard decoupled formulation is recovered. For ribbon models with bending-twisting coupling, the 
cross-derivative $\partial^2\mathcal{E}_k/\partial\kappa^{(2)}\partial\tau$ is nonzero and essential. The geometric stiffness term accounts for the nonlinearity of the strain-to-DOF mapping and takes the same form across all models.

For non-local models, differentiating Equation~\eqref{eq:globalGradient} fully would additionally yield off-diagonal Hessian blocks coupling DOF sets $\mathcal{I}_m$ and $\mathcal{I}_k$ for $m \neq k$. These off-diagonal contributions are omitted from the assembly, retaining the banded sparsity structure of $\mathbf{H}_E$ at negligible cost in accuracy for typical mesh resolutions.

\subsection{Temporal Integration}

Static equilibrium requires $\nabla_{\mathbf{q}} E = \mathbf{F}_{\text{ext}}$. However, solving this directly can fail to capture multistable configurations characteristic of ribbon mechanics. The inclusion of inertial terms regularizes the problem, enabling the solver to traverse energy barriers and converge to physically meaningful equilibria, particularly near instabilities and supercritical bifurcation points where multiple stable configurations coexist.

The equations of motion for the discrete ribbon take the standard second-order form
\begin{equation}
\mathbf{M}\ddot{\mathbf{q}} = \mathbf{F}_{\text{int}}(\mathbf{q}) + \mathbf{F}_{\text{ext}}(\mathbf{q}, t),
\end{equation}
where $\mathbf{M}$ is the lumped mass matrix, $\mathbf{F}_{\text{int}} = -\nabla_{\mathbf{q}} E$ is the internal elastic force, and $\mathbf{F}_{\text{ext}}$ are all external and boundary forces. We discretize this system using an implicit Euler scheme. Denoting the time step by $h$ and the displacement increment by $\Delta\mathbf{q}_{k+1} = \mathbf{q}_{k+1} - \mathbf{q}_k$, the implicit Euler update requires
\begin{equation}
\mathbf{M}\Delta\mathbf{q}_{k+1} - h\mathbf{M}\dot{\mathbf{q}}_k - h^2[\mathbf{F}_{\text{int},k+1} + \mathbf{F}_{\text{ext},k+1}] = \mathbf{0}.
\end{equation}

Newton-Raphson iteration requires the Jacobian matrix
\begin{equation}
\mathbf{J} = \mathbf{M} - h^2\left[\mathbf{H}_E + \frac{\partial\mathbf{F}_{\text{ext}}}{\partial\mathbf{q}}\right].
\end{equation}

An adaptive time stepping scheme balances computational efficiency with the accurate resolution of critical phenomena. The time step is reduced in our algorithm near bifurcation points and instabilities to capture the transition between stable branches, and increased during smooth deformation regimes to accelerate convergence (see Algorithm~\ref{alg:implicit_euler}). This adaptivity is essential for simulating ribbons, where snap-through buckling and bistable switching demand fine temporal resolution for convergence while quasi-static loading permits larger steps.

Algorithms~\ref{alg:implicit_euler} and~\ref{alg:robust_solve} together constitute the simulation solver. The simulation runs from $t = 0$ to a prescribed end time $T_{\text{end}}$ with time step $h$ bounded by $[h_{\min}, h_{\max}]$. At each step, Newton-Raphson iteration drives the free DOF residual $\mathbf{r}_{\text{free}}$ to zero, where ``free'' denotes the unconstrained DOFs in $\mathbf{q}$ after removing boundary condition DOFs. Convergence is declared when either the force residual norm $\|\mathbf{r}_{\text{free}}\|$ falls below $\delta_F$ (force tolerance) or the normalized displacement increment $\|\Delta\mathbf{q}_{\text{free}}\|_\infty / h$ falls below $\delta_u$ (displacement tolerance). Here the infinity norm $\|\cdot\|_\infty = \max_i |\cdot_i|$ takes the maximum absolute component, bounding the worst-case nodal displacement rather than the root mean square across all degrees of freedom, which is a more conservative convergence criterion. After each successful step, the material directors $\mathbf{m}_1^i, \mathbf{m}_2^i$ and the reference twist $m_{\text{ref}}^i$ are updated via parallel transport. The step size is reduced by a configurable factor ($0.5$ in our implementation) on convergence failure (floored at $h_{\min}$) and increased (with cap at $h_{\max}$) by a configurable factor ($1.5$ in our implementation) when the peak displacement $\|\Delta\mathbf{q}_{\text{free}}\|_\infty$ falls below a stability threshold $\delta_{\text{stable}}$ and the iteration count is below $N_{\text{stable}}$, indicating a smooth deformation regime. When the Newton-Raphson Jacobian $\mathbf{J}_{\text{free}}$ is ill-conditioned, the linear solve is delegated to a robust solver (Algorithm~\ref{alg:robust_solve}), which applies Tikhonov-Miller regularization $(\mathbf{J} + \lambda\mathbf{I})$ with $\lambda$ increased geometrically from an initial value $\lambda_0$ until the condition number falls below threshold $\mathcal{K}_{\max}$. If regularization alone is insufficient, a truncated SVD pseudo-inverse $\mathbf{J}^{+}$ serves as a last resort.

The banded structure of $\mathbf{J}$, inherited from the local support of strain measures, enables $\mathcal{O}(M)$ computational complexity per iteration. The availability of exact analytical gradients and Hessians of the strain measures with respect to the geometric DOFs, combined with autograd-computed derivatives of the energy with respect to strains, enables fully implicit integration within this generalized framework, regardless of the specific energy model.

\begin{algorithm}[t]
\caption{Adaptive Implicit Euler Integration}
\label{alg:implicit_euler}
\begin{algorithmic}[1]
\REQUIRE Initial state $\mathbf{q}_0$, $\dot{\mathbf{q}}_0$, total time $T_{\text{end}}$, time step bounds $h_{\min}, h_{\max}$, tolerances $\delta_F$ (force), $\delta_u$ (displacement), stability thresholds $\delta_{\text{stable}}$, $N_{\text{stable}}$
\ENSURE Equilibrium configuration $\mathbf{q}$
\WHILE{$t < T_{\text{end}}$}
    \STATE \textbf{Newton-Raphson iteration:}
    \REPEAT
        \STATE Compute $\mathbf{F}_{\text{int}} = -\nabla_{\mathbf{q}} E$, $\mathbf{H}_E = \nabla^2_{\mathbf{q}} E$ from generalized energy
        \STATE Assemble residual $\mathbf{r} = \mathbf{M}\Delta\mathbf{q} - h\mathbf{M}\dot{\mathbf{q}} - h^2(\mathbf{F}_{\text{int}} + \mathbf{F}_{\text{ext}})$
        \STATE Assemble Jacobian $\mathbf{J} = \mathbf{M} - h^2(\mathbf{H}_E + \partial\mathbf{F}_{\text{ext}}/\partial\mathbf{q})$
        \STATE $\Delta\mathbf{q}_{\text{free}} \leftarrow \textsc{RobustSolve}(\mathbf{J}_{\text{free}}, \mathbf{r}_{\text{free}})$ \COMMENT{Algorithm~\ref{alg:robust_solve}}
        \STATE $\mathbf{q}_{\text{free}} \leftarrow \mathbf{q}_{\text{free}} - \Delta\mathbf{q}_{\text{free}}$
    \UNTIL{$\|\mathbf{r}_{\text{free}}\| < \delta_F$ \textbf{or} $\|\Delta\mathbf{q}_{\text{free}}\|_\infty / h < \delta_u$}
    \STATE \textbf{Adaptive time-stepping:}
    \IF{not converged}
        \STATE $h \leftarrow \max(h/2,\, h_{\min})$; retry step
    \ELSIF{$\|\Delta\mathbf{q}_{\text{free}}\|_\infty < \delta_{\text{stable}}$ and iterations $< N_{\text{stable}}$}
        \STATE $h \leftarrow \min(1.5h,\, h_{\max})$ \COMMENT{Increase step}
    \ENDIF
    \STATE Update $\dot{\mathbf{q}},\, \mathbf{m}_1^i,\, \mathbf{m}_2^i,\, m_{\text{ref}}^i$;\quad $t \leftarrow t + h$
\ENDWHILE
\RETURN $\mathbf{q}$
\end{algorithmic}
\end{algorithm}

\begin{algorithm}[t]
\caption{Robust Linear Solve with Regularization}
\label{alg:robust_solve}
\begin{algorithmic}[1]
\REQUIRE Jacobian $\mathbf{J}$, residual $\mathbf{r}$, condition threshold $\mathcal{K}_{\max}$, initial regularization $\lambda_0$
\ENSURE Newton step $\Delta\mathbf{q}$
\IF{$\text{cond}(\mathbf{J}) < \mathcal{K}_{\max}$}
    \RETURN $\mathbf{J}^{-1}\mathbf{r}$ \COMMENT{Direct solve}
\ENDIF
\STATE $\lambda \leftarrow \lambda_0$
\WHILE{$\text{cond}(\mathbf{J} + \lambda\mathbf{I}) \geq \mathcal{K}_{\max}$}
    \STATE $\lambda \leftarrow 10\lambda$ \COMMENT{Tikhonov-Miller regularization}
\ENDWHILE
\IF{regularization sufficient}
    \RETURN $(\mathbf{J} + \lambda\mathbf{I})^{-1}\mathbf{r}$
\ELSE
    \RETURN $\mathbf{J}^{+}\mathbf{r}$ \COMMENT{Truncated SVD pseudo-inverse fallback}
\ENDIF
\end{algorithmic}
\end{algorithm}

\section{Discrete Elastic Ribbon Models}
\label{sec:models}

\subsection{Discretized Energy Functional}

We now formulate the discrete elastic potential energy corresponding to the continuous models summarized in Table~\ref{tab:ribbon_energies}. The discretization follows the framework established in Section~\ref{sec:kinematics}, where curvatures and twist are defined at internal nodes.

\subsubsection{Discrete Kirchhoff Energy}

The discrete Kirchhoff energy $E^{\text{Kirc}}$ directly discretizes the continuous functional from Table~\ref{tab:ribbon_energies}, with an additional stretching term to handle extensibility
\begin{equation}
E^{\text{Kirc}} = E_s + E_{b1} + E_{b2} + E_t.
\end{equation}

The stretching energy penalizes axial deformation
\begin{equation}
E_s = \frac{1}{2}\sum_{i=0}^{M-2} EA(\varepsilon_i)^2 |\overline{\mathbf{e}}_i|.
\end{equation}

The major-axis bending energy corresponds to the ``hard'' direction
\begin{equation}
E_{b1} = \frac{1}{2}\sum_{i=1}^{M-1} \frac{EI_1}{\Delta l_i}
\left(\kappa_i^{(1)} - \overline{\kappa}_i^{(1)} \right)^2.
\end{equation}

The minor-axis bending energy corresponds to the ``easy'' direction
\begin{equation}
E_{b2} = \frac{1}{2}\sum_{i=1}^{M-1} \frac{EI_2}{\Delta l_i}
\left(\kappa_i^{(2)} - \overline{\kappa}_i^{(2)} \right)^2.
\end{equation}

The twisting energy penalizes torsional deformation
\begin{equation}
E_t = \frac{1}{2}\sum_{i=1}^{M-1} \frac{GJ}{\Delta l_i} \left(\tau_i - \overline{\tau}_i \right)^2.
\end{equation}

Here $\overline{\kappa}_i^{(1)}$, $\overline{\kappa}_i^{(2)}$, and $\overline{\tau}_i$ denote the natural strains in the reference configuration for the major-axis curvature, minor-axis curvature, and twist, respectively. Together we can define the natural strain vector $\overline{\boldsymbol{\epsilon}}_k = [\overline{\kappa}^{(1)}, \overline{\kappa}^{(2)}, \overline{\tau}]_k^T$. For a ribbon fabricated stress-free and straight, all natural strains become zero whereas for pre-curved or pre-twisted geometries they encode the rest shape. The Voronoi length $\Delta l_i$ is the reference length assigned to node $i$. This formulation serves as the baseline for subsequent ribbon models.

\subsubsection{Discrete Wunderlich Energy}\label{sec:discrete-wunderlich}

Following the continuous formulation, we replace the decoupled minor-axis bending and twisting terms with a unified coupled term
\begin{equation}
E^{\text{Wund}} = E_s + E_{b1} + E^{\text{Wund}}_{b_2t},
\end{equation}
where
\begin{equation}
E^{\text{Wund}}_{b_2t} = \frac{1}{2}\sum_{i=1}^{M-1} \frac{EI_2}{\Delta l_i} \left[ \left((\kappa_i^{(2)} - \overline{\kappa}_i^{(2)})(1 + \eta_i^2)\right)^2 \frac{1}{W\eta'_i} \log\left(\frac{1 + W\eta'_i/2}{1 - W\eta'_i/2}\right) \right].
\end{equation}

Here, $\eta_i = (\tau_i - \overline{\tau}_i) / (\kappa_i^{(2)} - \overline{\kappa}_i^{(2)})$ is the discrete analogue of the continuous ratio. The derivative term $\eta'_i$ is computed using finite differences of these deformation variables across neighboring elements. For an internal node, the central difference approximation is given by
\begin{equation}
\eta'_i \approx \frac{\Delta l_i}{2} \frac{\Delta \left( \tau\kappa^{(2)} \right)}{\left( \kappa_i^{(2)} - \overline{\kappa}_i^{(2)} \right)^2},
\end{equation}
where the numerator $\Delta(\tau\kappa^{(2)})$ represents the discrete derivative of the coupled deformations
\begin{equation}
\Delta \left( \tau\kappa^{(2)} \right) = \left( \Delta\tau_{i+1} - \Delta\tau_{i-1} \right) \left( \Delta\kappa_i^{(2)} \right) - \Delta\tau_i \left( \Delta\kappa_{i+1}^{(2)} - \Delta\kappa_{i-1}^{(2)} \right),
\end{equation}
using the notation $\Delta\phi_k = \phi_k - \overline{\phi}_k$ for the deformation from the natural state. 

\subsubsection{Discrete Sadowsky Energy}

As the Sadowsky model augments the Kirchhoff energy with a coupling term (Table~\ref{tab:ribbon_energies}), we write
\begin{equation}
E^{\text{Sado}} = E^{\text{Kirc}} + E^{\text{Sado}}_{b_2t}.
\end{equation}
The nonlinear coupling term discretizes directly as
\begin{equation}
E^{\text{Sado}}_{b_2t} = \frac{1}{2}\sum_{i=1}^{M-1} \frac{EI_2}{\Delta l_i} \frac{\left(\tau_i - \overline{\tau}_i \right)^4}{ \left(\kappa_i^{(2)} - \overline{\kappa}_i^{(2)} \right)^2 + \epsilon}.
\end{equation}
A small regularization constant $\epsilon$ is included to prevent numerical divergence when $\kappa_i^{(2)} \to \overline{\kappa}_i^{(2)}$.

\subsubsection{Discrete Sano Energy}

The Sano regularization replaces the numerical constant $\epsilon$ with the physically motivated parameter $\zeta$ from Table~\ref{tab:ribbon_energies}
\begin{equation}
E^{\text{Sano}} = E^{\text{Kirc}} + E^{\text{Sano}}_{b_2t}
\end{equation}
\begin{equation}
E^{\text{Sano}}_{b_2t} = \frac{1}{2}\sum_{i=1}^{M-1} \frac{EI_2}{\Delta l_i} \frac{\left(\tau_i - \overline{\tau}_i\right)^4}{1/(\zeta/\Delta l_i)^2 + \left(\kappa_i^{(2)} - \overline{\kappa}_i^{(2)}\right)^2}.
\end{equation}
The scaling $\zeta/\Delta l_i$ ensures consistency across mesh resolutions, bounding the energy as $\kappa_i^{(2)} \to \overline{\kappa}_i^{(2)}$.

\subsubsection{Discrete Audoly Energy}

Following the continuous formulation, we augment the Kirchhoff energy with a correction term modulated by the transition function $\varphi$
\begin{equation}
E^{\text{Aud}} = E^{\text{Kirc}} + E^{\text{Aud}}_{b_2t},
\end{equation}
\begin{equation}
E^{\text{Aud}}_{b_2t} = 3\sum_{i=1}^{M-1} \frac{EI_2 W^4}{b^2 (\Delta l_i)^3} \left[ \nu \left(\kappa_i^{(2)} - \overline{\kappa}_i^{(2)}\right)^2 + \left(\tau_i - \overline{\tau}_i\right)^2 \right]^2 \varphi(v_i).
\end{equation}
The dimensionless strain parameter is
\begin{equation}
v_i = \sqrt{12(1-\nu^2)} \frac{W^2}{b} \frac{\left(\kappa_i^{(2)} - \overline{\kappa}_i^{(2)}\right)}{\Delta l_i}.
\end{equation}
And the transition function is
\begin{equation}
    \varphi(v) = \frac{4}{v^2} \left( \frac{1}{2} - \frac{\cosh\sqrt{\frac{|v|}{2}} - \cos\sqrt{\frac{|v|}{2}}}{\sqrt{\frac{|v|}{2}} \left(\sinh\sqrt{\frac{|v|}{2}} + \sin\sqrt{\frac{|v|}{2}}\right)} \right).
\end{equation}
This formulation recovers the discrete Kirchhoff energy when $|v_i| \ll 1$ and approaches the Sadowsky limit when $|v_i| \gg 1$. Setting $\nu = 0$ with an appropriate rational approximation for $\varphi$ recovers the Sano formulation.

\subsection{Implementation Details}

Algorithm~\ref{alg:general_energy} summarizes the element-level assembly of elastic forces and the stiffness matrix. Each element $k$ spans three consecutive nodes $(\mathbf{x}_{k-1}, \mathbf{x}_k, \mathbf{x}_{k+1})$ and two adjacent twist angles $(\theta^{k-1}, \theta^k)$, giving 11 local DOFs mapped to global indices via $\mathcal{I}_k$. The ribbon thickness $b$ and Voronoi length $\Delta l_k$ together define the normalization of the curvature and twist components of $\hat{\boldsymbol{\epsilon}}_k$, which is the dimensionless strain vector passed to the energy model. Steps~1 and~3 use exact analytical expressions for the strain Jacobians $\mathbf{G}_k$ and Hessians $\mathbf{H}_{\epsilon,k}$ with respect to the local DOFs, derived in ~\ref{app:strain_gradients} - \ref{app:hessian_derivation}. Step~2 evaluates the scalar energy $\mathcal{E}_k$ and its derivatives with respect to $\hat{\boldsymbol{\epsilon}}_j$ for all $j$ on which $\mathcal{E}_k$ depends, using PyTorch autograd. For energy models where $\mathcal{E}_k$ depends only on the strains of element $k$ itself, the scatter in Step~4 reduces to $\mathcal{I}_k$ alone. For models where the energy at $k$ depends on strains of neighboring elements (e.g. Wunderlich energy), the gradient contributions are additionally scattered to their corresponding global DOF indices.

The key feature of this procedure is the clean separation between geometry and constitutive modeling. Steps~1 and~3 involve only the discrete differential geometry of the centerline, which are identical for all five energy models, while Step~2 queries the energy function purely in strain space. This modular structure means that implementing a new ribbon energy model requires only specifying the scalar function $\mathcal{E}(\hat{\boldsymbol{\epsilon}})$, with no modification to the geometric framework or the global assembly logic. While developed and applied here in the context of elastic ribbons, the framework is not specific to ribbons, it can potentially be extended to any one-dimensional energy model whose elastic potential depends on coupled strain measures, including models for elastoplastic rods~\cite{li2020discrete, Herrnboeck2021}, or other slender structures with nonstandard constitutive laws.

\begin{algorithm}[htbp]
\caption{Generalized Elastic Energy Gradient and Hessian Assembly}
\label{alg:general_energy}
\begin{algorithmic}
\REQUIRE Current state $\mathbf{q}$, material frames $\{\mathbf{m}_1^i, \mathbf{m}_2^i\}$,
         natural strains $\{\overline{\boldsymbol{\epsilon}}_k\}$, energy model 
         $\mathcal{E}_k(\{\boldsymbol{\epsilon}_j\}_{j \in \mathcal{N}(k)})$
\REQUIRE Global DOF index map $\mathcal{I}_k$: 11 local DOFs of element $k$ 
         ($\mathbf{x}_{k-1}, \mathbf{x}_k, \mathbf{x}_{k+1}, \theta^{k-1}, \theta^k$)
\ENSURE Global force vector $\mathbf{F}$, stiffness matrix $\mathbf{K}$
\STATE Initialize $\mathbf{F} \leftarrow \mathbf{0}$, $\mathbf{K} \leftarrow \mathbf{0}$
\STATE \textbf{Step 1: Strains and geometric derivatives (analytical, batched)}
\FOR{each element $k$}
    \STATE Compute $\boldsymbol{\epsilon}_k$, $\mathbf{G}_k = \partial\boldsymbol{\epsilon}_k/\partial\mathbf{q}[\mathcal{I}_k]$, $\mathbf{H}_{\epsilon,k} = \partial^2\boldsymbol{\epsilon}_k/\partial\mathbf{q}[\mathcal{I}_k]^2$ \COMMENT{~\ref{app:strain_gradients}}
    \STATE $\Delta\boldsymbol{\epsilon}_k \leftarrow \boldsymbol{\epsilon}_k - \overline{\boldsymbol{\epsilon}}_k$
\ENDFOR
\STATE \textbf{Step 2: Energy derivatives in strain space (autograd, batched)}
\STATE For each $k$, evaluate $\mathcal{E}_k$ and $\partial\mathcal{E}_k/\partial\boldsymbol{\epsilon}_{j}$ for all $j \in \mathcal{N}(k)$
\STATE \textbf{Step 3: Chain rule to local DOFs stencil (batched)}
\FOR{each element $k$}
    \STATE $\nabla_{\mathbf{q}[\mathcal{I}_k]} E_k \leftarrow \mathbf{G}_k^T\, \partial\mathcal{E}_k/\partial\boldsymbol{\epsilon}_k \;+\; \sum_{l} (\partial\mathcal{E}_k/\partial\epsilon_{k,l})\,[\mathbf{H}_{\epsilon,k}]_l$
    \STATE $\nabla^2_{\mathbf{q}[\mathcal{I}_k]} E_k \leftarrow \mathbf{G}_k^T\,(\nabla^2_{\boldsymbol{\epsilon}_k}\mathcal{E}_k)\,\mathbf{G}_k \;+\; \sum_{l} (\partial\mathcal{E}_k/\partial\epsilon_{k,l})\,[\mathbf{H}_{\epsilon,k}]_l$
\ENDFOR
\STATE \textbf{Step 4: Scatter to global DOFs stencil}
\FOR{each element $k$}
    \FOR{each $j \in \mathcal{N}(k)$ such that $\partial\mathcal{E}_k/\partial\boldsymbol{\epsilon}_j \neq \mathbf{0}$}
        \STATE $\mathbf{F}[\mathcal{I}_j] \leftarrow \mathbf{F}[\mathcal{I}_j] - \mathbf{G}_j^T\,\partial\mathcal{E}_k/\partial\boldsymbol{\epsilon}_j$
    \ENDFOR
    \STATE $\mathbf{K}[\mathcal{I}_k,\mathcal{I}_k] \leftarrow \mathbf{K}[\mathcal{I}_k,\mathcal{I}_k] - \nabla^2_{\mathbf{q}[\mathcal{I}_k]} E_k$
\ENDFOR
\end{algorithmic}
\end{algorithm}

\section{Shear Induced Bifurcation}\label{sec:results}

\subsection{Overview}

Inspired by the experimental and computational study of Huang et al.\ \cite{huang2020shear}, we adopt the same boundary value problem that isolates the shear-induced supercritical pitchfork bifurcation of a pre-buckled elastic ribbon. A ribbon of length $L = 0.1$~m, width $W$, and thickness $b$ is clamped at both ends with the clamp faces oriented parallel to the width direction. The clamp angle is held fixed at $\psi = 0^\circ$ throughout the loading sequence. A longitudinal compression of $\Delta L/L = 1/4$ is first applied quasi-statically by displacing one clamped end toward the other, maintaining a fixed span of $L - \Delta L = 75$~mm. This compression exceeds the Euler buckling threshold and drives the ribbon into a buckled equilibrium (Figure~\ref{fig:overview}a). Once this pre-buckled configuration is established, a transverse (lateral) displacement $\Delta W$ is imposed incrementally at one clamped end to induce shear (Figure~\ref{fig:overview}b). The material is taken to be linearly elastic with Young's modulus $Y = 10$~GPa and Poisson's ratio $\nu = 0.5$. Since the problem is governed by geometry rather than absolute stiffness, the specific value of the modulus does not influence the normalized results but sets the force scale.

We consider two values of the ribbon thickness: $b = 0.1$~mm and $b = 1$~mm, corresponding to length-to-thickness ratios of $L/b = 1000$ and $L/b = 100$, respectively. Although Huang et al.\ \cite{huang2020shear} systematically varied the width-to-length ratio $W/L$ and the pre-compression $\Delta L/L$, the effect of thickness on the bifurcation response within the one-dimensional modeling framework was not investigated. Our study addresses this gap by examining how the thickness parameter, which enters the various ribbon energy models through distinct mechanisms: (i) the regularization parameter $\zeta$ in the Sano model, (ii) the characteristic curvature $\kappa^*$ in the Audoly model, and (iii) the bending-to-stretching stiffness ratio in the plate model influences the predicted bifurcation thresholds and post-bifurcation behavior.

Based on previous experimental and computational observations \cite{yu2019bifurcations,huang2020shear}, the pre-buckled ribbon is expected to undergo a well-defined sequence of configuration changes as the normalized transverse shear $\Delta \bar{W} = \Delta W/L$ increases. The initial buckled state is a symmetric $U$-shaped configuration in which the ribbon arches upward (or downward) with no twist. At a first critical shear $\Delta \bar{W}_1$, a supercritical pitchfork bifurcation occurs: the symmetric $U$ configuration becomes unstable and the ribbon transitions into one of two symmetric $US$ configurations ($US+$ or $US-$), characterized by a combination of bending and localized twist. At a second threshold $\Delta \bar{W}_2$, the ribbon undergoes a further transition into a fully twisted $S$-shaped configuration. The two-dimensional Discrete Elastic Plates (DEP) model used by Huang et al.\ \cite{huang2020shear}, which converges to the F\"{o}ppl-von K\'{a}rm\'{a}n equations, accurately reproduces these transitions across a wide range of width-to-length ratios and matches experimental observations. The one-dimensional anisotropic Kirchhoff rod model, by contrast, predicts bifurcation thresholds that are independent of the ribbon width, yielding accurate results only for narrow strips ($W/L \lesssim 1/20$) and progressively overestimating the critical shear as the width increases.

The developable ribbon models---Sadowsky \& Wunderlich---are expected to face a more fundamental difficulty with this benchmark problem. As discussed by Yu \& Hanna \cite{yu2019bifurcations} and confirmed computationally for Sadowsky by Huang et al.\ \cite{huang2020shear}, these models enforce vanishing Gaussian curvature everywhere on the ribbon surface. At inflection points, where the bending curvature $\kappa^{(2)}$ passes through zero, the developability constraint forces the twist $\tau$ to vanish simultaneously, creating a discontinuity in the curvature profile and an associated energy barrier. This barrier prevents the ribbon from smoothly transitioning between configurations that differ in the number of inflection points. Since the $U \to US$ transition requires the spontaneous creation of new inflection points, a Sadowsky or Wunderlich simulation initialized from the symmetric $U$ state cannot discover the bifurcation by itself, so it remains trapped in the $U$ branch even beyond the physical critical shear. These models can reproduce the $US$ and $S$ configurations only when seeded with an appropriate initial guess obtained from a rod or plate simulation \cite{huang2020shear}. This limitation motivates the inclusion of the regularized models of Sano and Audoly, which relax the strict developability constraint and may permit the bifurcation to occur spontaneously.

At the supercritical pitchfork point, the ribbon may deflect into either the $US+$ or $US-$ branch, with the selection governed by infinitesimal perturbations. In principle, the response under positive and negative transverse displacement should be symmetric, producing identical force-displacement curves with no dependence on the direction of shear. In practice, however, the bifurcation is sensitive to small perturbations accumulated during the quasi-static loading history. We occasionally observe that, during positive transverse displacement, the solver snaps from one symmetric $US$ branch onto the other (e.g., $US- \to US+$), while the negative transverse direction traverses the same shear range without such a branch switch. This asymmetry is a numerical artifact of the path dependence near the pitchfork point rather than a physical feature of the problem. In such cases, we select the negative transverse direction, which remains on a single $US$ branch throughout and exhibits a smoothly varying total elastic energy, ensuring a consistent and unbiased comparison across all energy models and discretizations.

To provide a model-independent reference against which all five one-dimensional energy formulations can be assessed, we compare our results with finite element analysis (FEA) performed using shell elements. The FEA solution, which resolves both bending and mid-surface stretching without any dimensional reduction or developability assumption, serves as a surrogate for the full two-dimensional plate mechanics. This choice is motivated by the observation of Huang et al.\ \cite{huang2020shear} that the DEP plate model matches experimental data to within measurement uncertainty across all tested geometries, establishing it as a reliable ground truth. By benchmarking all five one-dimensional ribbon models against a common FEA reference under identical boundary conditions, we ensure that any differences in the predicted response can be attributed directly to the constitutive assumptions of each energy formulation rather than to inconsistencies in numerical implementation.

\subsection{Comparing all Energy Models}

We assess the five one-dimensional ribbon energy models by plotting the normalized midpoint height $\bar{H}_m = |H_m|/L$ as a function of the normalized transverse shear $\Delta \bar{W} = \Delta W/L$, and comparing against shell-element FEA results across varying width-to-length ratios $W/L \in \{1/2, 1/6, 1/12, 1/20\}$. The absolute value of midpoint height is taken because the direction of initial buckling, upward or downward, is a stochastic symmetry breaking that depends on numerical perturbations, and the sign of $H_m$ carries no significance for this comparison.

\subsubsection{Thin ribbon: $L/b = 1000$}

\begin{figure}[htbp]
    \centering
    \includegraphics[width=1.0\textwidth]{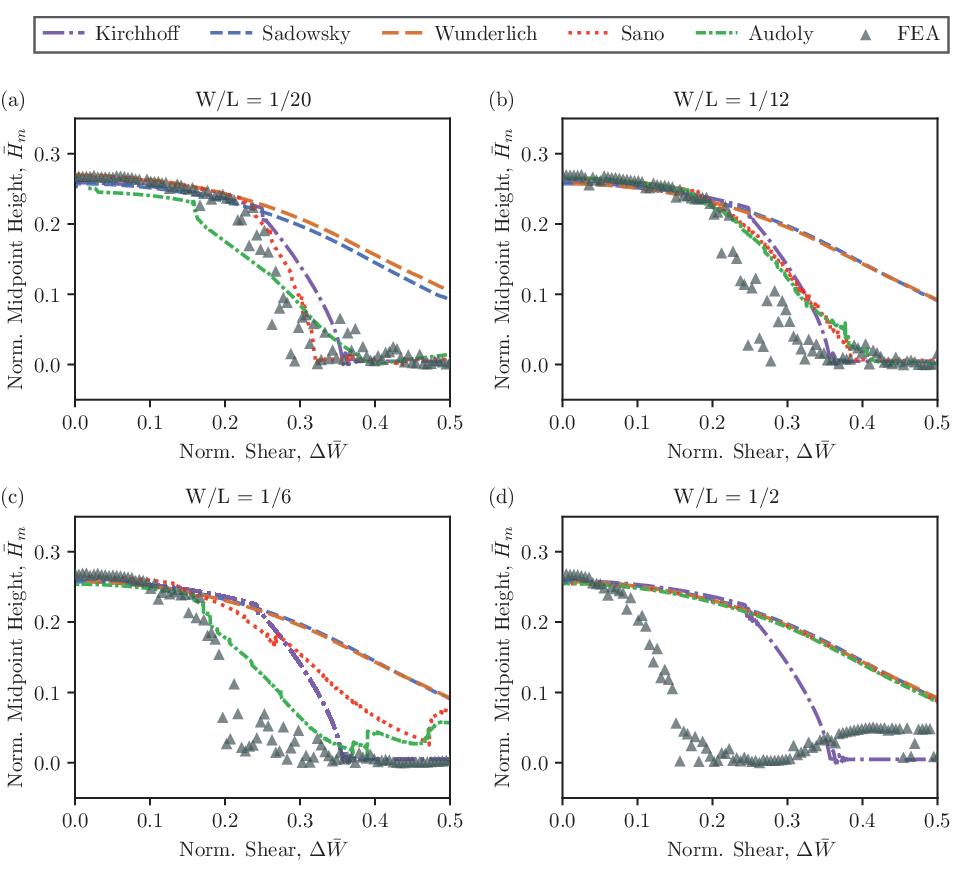}
    \caption{Comparing all centerline-based energy models against FEA for shear-induced bifurcation, at different widths $W/L \in \{ 1/20, 1/12, 1/6, 1/2 \}$ for $L/b=1000$.}
    \label{fig:all-energies-hmid-b1e-4}
\end{figure}

We first examine the case $b = 0.1$~mm ($L/b = 1000$), shown in Figure~\ref{fig:all-energies-hmid-b1e-4}(a-d) at four different widths $W/L \in \{ 1/20, 1/12, 1/6, 1/2 \}$. The FEA reference captures the expected phenomenology that the supercritical pitchfork point shifts to lower values of $\Delta \bar{W}$ as the ribbon width increases, consistent with the findings of Huang et al.\ \cite{huang2020shear}. However, the FEA response at this thickness exhibits considerable non-smoothness in the vicinity of both the $U \to US$ and $US \to S$ transitions, reflecting the high stiffness of the nearly inextensible thin plate. We also performed DEP simulations at $L/b = 1000$ and encountered pronounced stiffness in the implicit Euler integration, necessitating aggressive adaptive time-stepping (Algorithm~\ref{alg:implicit_euler}). After applying adaptive refinement, the DEP results at $L/b = 1000$ diverge from those reported by Huang et al.\ \cite{huang2020shear}, whereas the DEP results at $L/b = 100$ reproduce their published curves. This discrepancy suggests that Huang et al.\ assumed thickness to have negligible influence due to inextensibility and reported results computed at $L/b \approx 100$ while comparing against experiments conducted at $L/b \approx 1000$. Our FEA and plate simulations confirm that, although the correct bifurcation trend is captured at both thickness values, the response is noticeably less smooth at the lower thickness.

Turning to the one-dimensional models, the Sadowsky and Wunderlich formulations fail to capture the supercritical pitchfork bifurcation at any width, remaining on the symmetric $U$ branch throughout the shear range (Figure~\ref{fig:all-energies-hmid-b1e-4}(a-d)). This is consistent with the energy barrier imposed by the strict developability constraint at inflection points, as discussed in the preceding section. The Kirchhoff model reproduces the bifurcation behavior reported by Huang et al.\ \cite{huang2020shear}: it yields reasonably accurate predictions at $W/L = 1/20$ (Figure~\ref{fig:all-energies-hmid-b1e-4}a), but the predicted bifurcation threshold remains invariant as the width increases (Figure~\ref{fig:all-energies-hmid-b1e-4}(b-d)), failing to capture the leftward shift observed in FEA. The Sano and Audoly models, exhibit qualitatively different behavior. At $W/L = 1/20$, the Sano model provides the closest agreement with the FEA reference among all one-dimensional models, outperforming both Kirchhoff and Audoly (Figure~\ref{fig:all-energies-hmid-b1e-4}a). The Audoly model at this width appears to lie between the Kirchhoff and Sadowsky responses, consistent with its role as a constitutive interpolant between the two limits, and corroborating the findings of Huang et al.\ \cite{huang2022discrete}, who employed a simplified variant of the Audoly-Neukirch formulation. As the width increases to $W/L = 1/12$, both the Sano and Audoly models correctly shift the $U \to US$ bifurcation point to lower shear values, matching the FEA trend  (Figure~\ref{fig:all-energies-hmid-b1e-4}b). However, both models predict a rightward shift of the $US \to S$ transition, which appears inaccurate relative to FEA. At $W/L = 1/6$, Audoly shows the closest overall agreement with the FEA response, though the $US \to S$ transition and post-transition behavior remain inaccurate  (Figure~\ref{fig:all-energies-hmid-b1e-4}c). At $W/L = 1/2$, both Sano and Audoly fail to capture the bifurcation branch entirely  (Figure~\ref{fig:all-energies-hmid-b1e-4}d).

\subsubsection{Thick ribbon: $L/b = 100$}

\begin{figure}[htbp]
    \centering
    \includegraphics[width=1.0\textwidth]{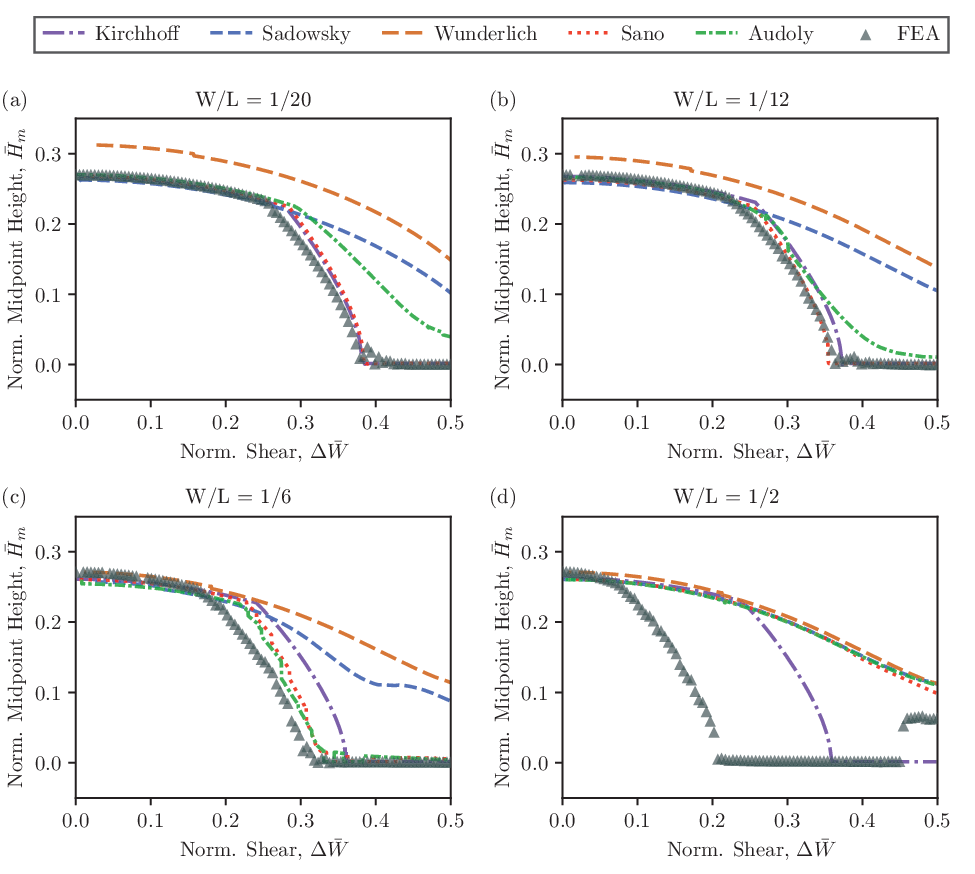}
    \caption{Comparing all 1D energy models against FEA for shear-induced bifurcation, at different ribbon widths $W/L \in \{ 1/20, 1/12, 1/6, 1/2 \}$ for $L/b=100$.}
    \label{fig:all-energies-hmid-b1e-3}
\end{figure}

Recognizing the non-smooth character of the FEA response at $L/b = 1000$, to establish a cleaner comparison, we next examine the case $b = 1$~mm ($L/b = 100$) in Figure~\ref{fig:all-energies-hmid-b1e-3}(a-d), again across the same four ribbon widths $W/L \in \{ 1/20, 1/12, 1/6, 1/2 \}$. As we can see in Figure~\ref{fig:all-energies-hmid-b1e-3}(a-d), both the FEA reference and the one-dimensional models exhibit comparatively smoother behavior at this thickness. The Wunderlich and Sadowsky models again fail to capture the bifurcation for all widths, confirming that the energy barrier at inflection points persists irrespective of the thickness parameter.

At $W/L = 1/20$, both the Sano and Kirchhoff models produce accurate bifurcation predictions relative to FEA (Figure~\ref{fig:all-energies-hmid-b1e-3}a). As the width increases to $W/L = 1/12$ and $W/L = 1/6$, a clear distinction emerges: the Sano model shifts both the $U \to US$ and $US \to S$ transition points to lower shear values with increasing width, whereas the Kirchhoff model remains unchanged (Figure~\ref{fig:all-energies-hmid-b1e-3}(b-c)). This contrasts with the $L/b = 1000$ case, where the Sano model captured the leftward shift of the first transition but incorrectly predicted a rightward shift of the second; at $L/b = 100$, both transitions move consistently leftward, in agreement with the FEA trend. The Sano model provides a reasonable quantitative match with FEA at $W/L = 1/12$, although at $W/L = 1/6$ (see Figure~\ref{fig:all-energies-hmid-b1e-3}c) the FEA transition points occur at even lower shear values than the Sano predictions. The Audoly model exhibits a markedly different behavior at this thickness. At $W/L = 1/20$, it again sits between the Kirchhoff and Sadowsky responses (Figure~\ref{fig:all-energies-hmid-b1e-3}a), reflecting its role as a constitutive interpolant between the two limits, corroborating the observations from the $L/b = 1000$ case and the findings of Huang et al.\ \cite{huang2022discrete}. At $W/L = 1/12$, this interpolant character persists, though the Audoly response shifts noticeably closer to Sano rather than Kirchhoff (Figure~\ref{fig:all-energies-hmid-b1e-3}b); despite this, the $US \to S$ transition remains significantly inaccurate relative to FEA. At $W/L = 1/6$, Audoly recovers and follows a response almost identical to that of Sano (Figure~\ref{fig:all-energies-hmid-b1e-3}c). At $W/L = 1/2$, both Sano and Audoly again fail to capture the bifurcation branch (Figure~\ref{fig:all-energies-hmid-b1e-3}d). Renderings of the deformation sequences for the Sano model across $W/L \in \{1/20, 1/12, 1/6, 1/2\}$ are provided in Supplementary Videos S1-S4, and those for the Audoly model across $W/L \in \{1/20, 1/12, 1/6\}$ in Supplementary Videos S7-S9.

\subsubsection{Summary}

Across both thickness values and widths spanning $W/L \in \{1/20, 1/12, 1/6\}$, the Sano model emerges as the most accurate one-dimensional formulation for this benchmark problem. It is the only model that consistently captures the leftward shift of the bifurcation threshold with increasing width while maintaining relatively closer quantitative agreement with the FEA reference. For the Audoly model, we can generalize that it consistently sits between the Sadowsky and the Sano (instead of Kirchhoff) across all conditions tested, lying closer to Sadowsky at narrow widths and progressively shifting toward Sano as the width increases. This interpolant behavior limits its accuracy at narrow widths where it under-performs relative to Sano \& FEA, particularly in resolving the $US \to S$ transition. The Kirchhoff model remains a reliable baseline for narrow ribbons but is fundamentally unable to account for width effects. The strict developable models, Sadowsky and Wunderlich, cannot spontaneously traverse the pitchfork bifurcation under any conditions tested. We quantify these observations more precisely in the following section.

\subsection{Analyzing Shear Force}

To move beyond the qualitative observations of the preceding section, we examine the normalized shear force, $\bar{F}_{shear} = F_{shear}L/(Yb^3)$, alongside the normalized midpoint height, $\bar{H}_m$, for the two interesting one-dimensional models at $L/b = 100$. The shear force provides a more sensitive diagnostic of the bifurcation than the midpoint displacement alone: the supercritical pitchfork manifests as a distinct peak in the force-displacement curve, enabling precise identification of the critical shear $\Delta W_1/L$ and the subsequent $US \to S$ transition at $\Delta W_2/L$.

\begin{figure}[hbtp]
    \centering
    \includegraphics[width=1.0\textwidth]{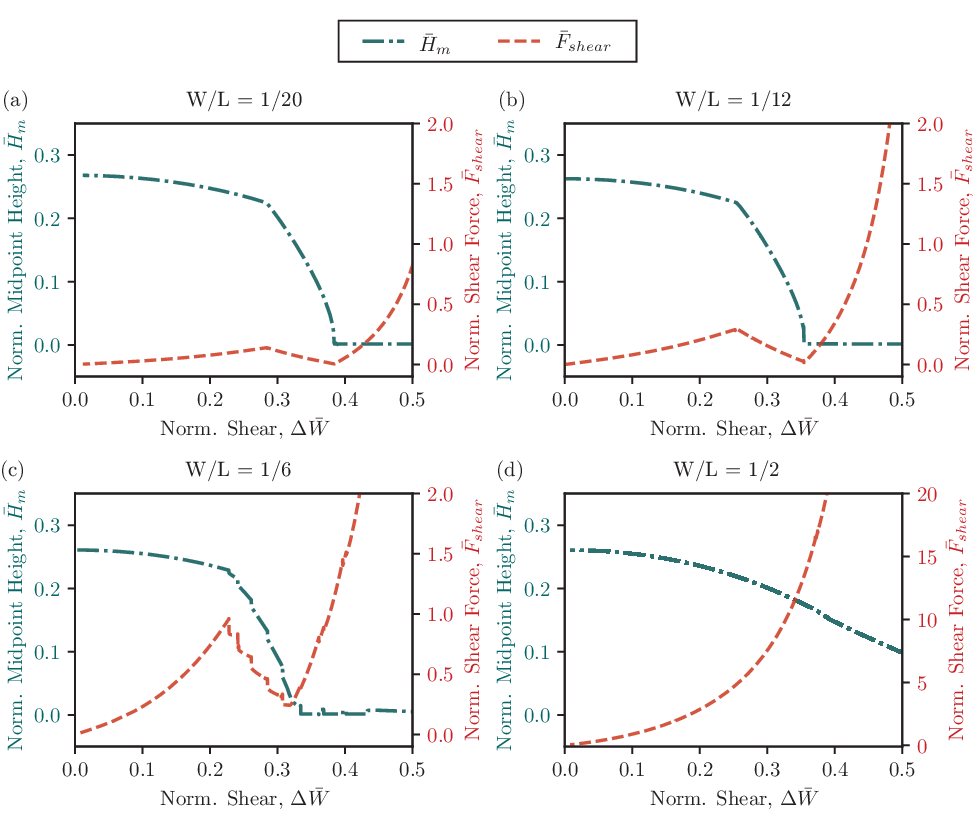}
    \caption{For Sano energy model: comparing normalized external shear force $\bar{F}_{shear} = F_{shear}L/(Yb^3)$ as functions of normalized transverse shear $\Delta \bar{W}$ at ribbon widths $W/L \in \{ 1/20, 1/12, 1/6, 1/2 \}$ for $L/b=100$.}
    \label{fig:shear-force-sano-b1e-3}
\end{figure}

For the Sano model (Figure~\ref{fig:shear-force-sano-b1e-3}(a-c)), the force-displacement curves exhibit a well-defined peak at the $U \to US$ transition and a subsequent local minimum at the $US \to S$ transition across $W/L \in \{1/20, 1/12, 1/6\}$. Two clear trends emerge as the width increases: the peak shear force shifts to lower values of $\Delta \bar{W}$, confirming the leftward migration of the bifurcation threshold, and the magnitude of the peak force itself increases monotonically with width. Both observations are consistent with the plate model results reported by Huang et al.\ \cite{huang2020shear}, where wider plates require less shear to bifurcate but develop larger internal forces due to the increased bending stiffness of the wider cross-section.

\begin{figure}[htbp]
    \centering
    \includegraphics[width=1.0\textwidth]{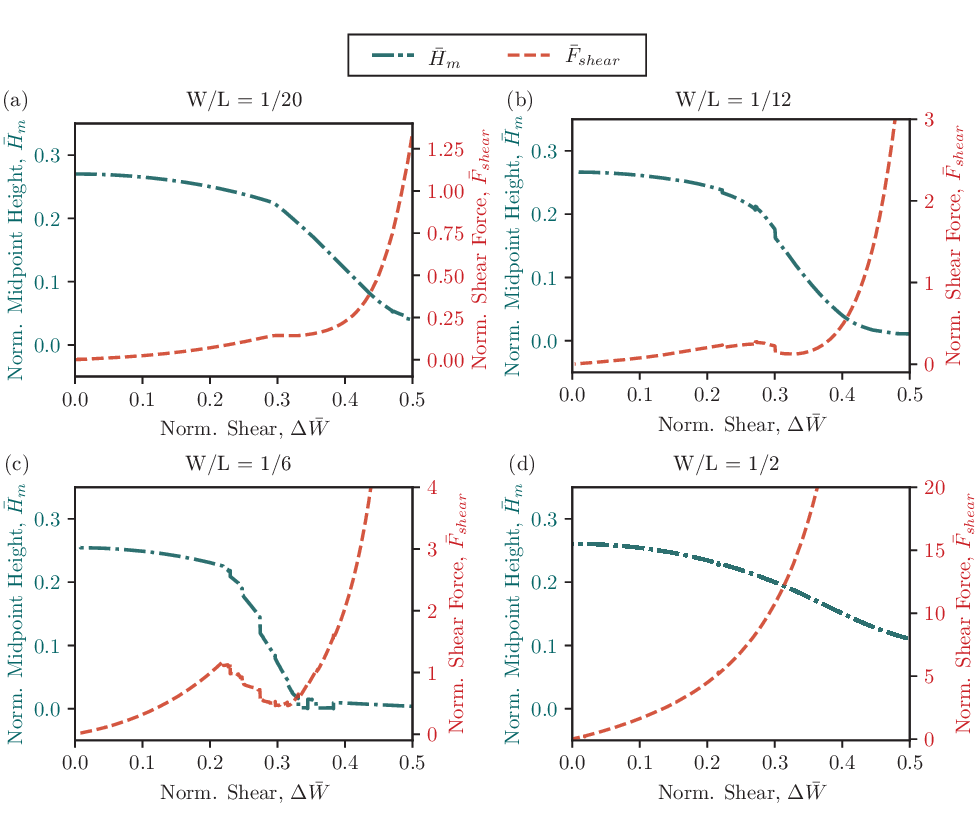}
    \caption{For Audoly energy model: comparing normalized external shear force $\bar{F}_{shear} = F_{shear}L/(Yb^3)$ as functions of normalized transverse shear $\Delta \bar{W}$ at ribbon widths $W/L \in \{ 1/20, 1/12, 1/6, 1/2 \}$ for $L/b=100$.}
    \label{fig:shear-force-audoly-b1e-3}
\end{figure}

For the Audoly model (Figure~\ref{fig:shear-force-audoly-b1e-3}(a-c)), the $U \to US$ bifurcation peak is deducible for all widths up to $W/L = 1/6$. However, the $US \to S$ transition cannot be clearly identified from the shear force curves at $W/L = 1/20$ (Figure~\ref{fig:shear-force-audoly-b1e-3}a) and $W/L = 1/12$ (Figure~\ref{fig:shear-force-audoly-b1e-3}b), corroborating the observation from the midpoint height plots that the Audoly model does not accurately resolve the second transition at these geometries. This difficulty likely stems from the behavior of the transition function $\varphi(v)$, which might stiffen the energy landscape against the torsional rearrangement required for the $US \to S$ transition, effectively raising the energy barrier between the two configurations.

At $W/L = 1/2$, both the Sano and Audoly models exhibit a qualitatively different response (Figure~\ref{fig:shear-force-sano-b1e-3}d, Figure~\ref{fig:shear-force-audoly-b1e-3}d). The normalized shear force rises to substantially larger magnitudes than at narrower widths, yet neither model undergoes the pitchfork bifurcation within the admissible shear range. We hypothesize that this failure is rooted in the strengthening of the developability-like constraint as the width approaches the length scale. In the Sano model, the regularization parameter $\zeta^2 = (1 - \nu)W^4/(60b^2)$ grows as $W^4$, causing the coupling term $\tau^4/(1/\zeta^2 + (\kappa^{(2)})^2)$ to impose an increasingly stiff penalty against configurations in which twist and curvature are not appropriately balanced. Similarly, in the Audoly model, the characteristic curvature $\kappa^* = b/(\sqrt{12(1-\nu^2)}\,W^2)$ decreases as $W^{-2}$, so that even modest bending curvatures drive the dimensionless strain parameter $v = \kappa^{(2)}/\kappa^*$ into the large-$v$ regime where the model approaches the Sadowsky limit. In both cases, the effective energy barrier at inflection points grows with width, and at $W/L = 1/2$ this barrier is sufficiently large to prevent the solver from discovering the bifurcated branch.

To quantify the accuracy of each model in predicting the bifurcation thresholds, we report the critical shear values at both the $U \to US$ and $US \to S$ transitions in Table~\ref{tab:bifurcation_shifts}. Taking the critical shear at $W/L = 1/20$ as the baseline for each model, we compute the percentage shift $\Delta_\%$ in the critical shear as the width increases to $W/L = 1/12$ and $W/L = 1/6$. We additionally report the absolute error $\delta_e$ of each one-dimensional model relative to the FEA reference at each width. The negative percentage shift $-\Delta_\%$ indicates that the bifurcation threshold has moved to lower shear values, which is the physically correct trend established by the FEA and plate simulations.

\begin{table}[htbp]
\centering
\small
\renewcommand{\arraystretch}{1.3}
\caption{Negative percentage shift $-\Delta_\%$ in critical shear from their respective $W/L = 1/20$ baselines and absolute error $\delta_e$ relative to FEA, for the $U \to US$ and $US \to S$ transitions at $L/b = 100$. An asterisk ($*$) denotes the FEA reference against which errors are computed. A dash (---) indicates that the transition was not clearly identifiable.}
\label{tab:bifurcation_shifts}
\begin{tabular}{@{} l l r rr rr @{}}
\toprule
& & \multicolumn{1}{c}{$W/L = 1/20$} & \multicolumn{2}{c}{$W/L = 1/12$} & \multicolumn{2}{c}{$W/L = 1/6$} \\
\cmidrule(lr){3-3} \cmidrule(lr){4-5} \cmidrule(lr){6-7}
Transition & Model & $\delta_e$ & $-\Delta_\%$ (\%) & $\delta_e$ & $-\Delta_\%$ (\%) & $\delta_e$ \\
\midrule
\multirow{4}{*}{$U \to US$}
 & FEA (GT)    & $*$  & 11.1 & $*$   & 33.3 & $*$   \\
 & Kirchhoff & \textbf{0.01} & 0.0  & 0.04 & 0.0  & 0.10 \\
 & Sano     & \textbf{0.01} & \textbf{8.9} & \textbf{0.005} & \underline{21.4} & \textbf{0.04} \\
 & Audoly   & \underline{0.025} & \underline{8.5}  & \underline{0.03} & \textbf{25.4} & \textbf{0.04} \\
\midrule
\multirow{4}{*}{$US \to S$}
 & FEA (GT)     & $*$  & 7.8 & $*$   & 18.4 & $*$   \\
 & Kirchhoff & \textbf{0.0} & 0.0  & \underline{0.03} &  0.0  & 0.07 \\
 & Sano     & \textbf{0.0} & \textbf{7.8}  & \textbf{0.0} & \textbf{17.1} & \textbf{0.005} \\
 & Audoly   & --- & ---     & ---   & --- & \textbf{0.005} \\
\bottomrule
\end{tabular}
\end{table}

Several observations follow from Table~\ref{tab:bifurcation_shifts}. First, the Kirchhoff model exhibits zero percentage shift by construction, since it contains no width dependence; its absolute error relative to FEA grows steadily with width, confirming that it becomes progressively less reliable for wider ribbons. Second, the Sano model consistently captures the largest fraction of the FEA percentage shift at both transitions and across both widths, yielding the smallest absolute errors among all one-dimensional models. Third, the Audoly model captures the correct trend for the $U \to US$ transition but, as noted above, fails to resolve the $US \to S$ transition at $W/L = 1/20$ and $W/L = 1/12$, limiting its utility as a general-purpose ribbon model for this class of problems. These quantitative results reinforce the conclusion that, among the one-dimensional formulations tested, the Sano model offers the best balance of physical fidelity and numerical robustness for predicting shear-induced bifurcations of pre-buckled ribbons.

\subsection{Homotopy-based Width Transition}

The analysis of the preceding sections established that the Sano model captures the width-dependent leftward shift of the bifurcation thresholds for $W/L \in \{1/20, 1/12, 1/6\}$, but fails to discover the bifurcated branch at $W/L = 1/2$ when the transverse shear is applied directly from the symmetric $U$ configuration. The question naturally arises: does a bifurcation branch exist at these larger widths within the Sano energy landscape, or does the model fundamentally preclude it?

To answer this question, we exploit the robustness of our Discrete Elastic Ribbon framework and introduce a homotopy-based loading strategy that circumvents the energy barrier preventing direct bifurcation at large widths. The procedure consists of three stages. First, we perform the standard quasi-static shear loading in the positive transverse direction using a narrower ribbon at $W/L = 1/12$, for which the solver reliably traverses the pitchfork bifurcation and reaches the $S$ branch. Second, while holding the transverse shear fixed, we homotopically increase the ribbon width from $W/L = 1/12$ to the target width ($W/L = 1/3$ or $W/L = 1/2$), allowing the ribbon to equilibrate at each intermediate width. This continuous deformation in the width parameter carries the solution along the bifurcated energy branch without requiring the solver to cross the energy barrier from the symmetric side. Third, with the ribbon now on the bifurcated branch at the target width, we reverse the transverse shear direction hence mapping out the full bifurcation curve. Corresponding renderings of this homotopy-based width transition for the Sano model at $W/L = 1/3$ and $W/L = 1/2$ are provided in Supplementary Videos S5 \& S6, respectively.

\begin{figure}[h]
    \centering
    \includegraphics[width=1.0\textwidth]{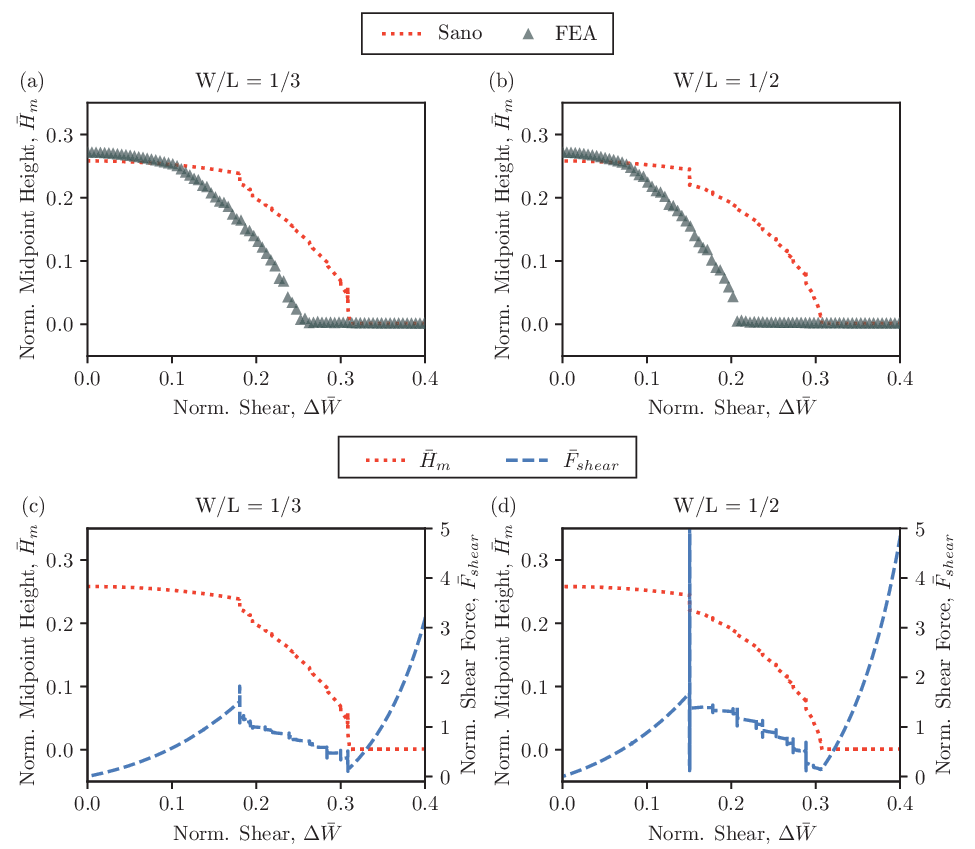}
    \caption{Sano's bifurcation behavior on quasi-static loading after homotopy-based width transition evaluated on ribbon widths $W/L \in \{ 1/3,  1/2 \}$ and $L/b=100$.}
    \label{fig:homotopy-width-sano}
\end{figure}

This strategy successfully reveals the bifurcated branches at both $W/L = 1/3$ and $W/L = 1/2$ within the Sano model, see Figure~\ref{fig:homotopy-width-sano}(a-d). The discovered branches exhibit the same qualitative structure, $U$, $US$, and $S$ configurations connected by transitions, as those found at narrower widths through direct shear loading. The quantitative results are summarized in Table~\ref{tab:bifurcation_shifts_homotopy}. For the $U \to US$ transition, the trend observed at smaller widths persists: the critical shear continues to shift leftward as the width increases from $W/L = 1/6$ through $W/L = 1/3$ (Figure~\ref{fig:homotopy-width-sano}(a,c)) to $W/L = 1/2$ (Figure~\ref{fig:homotopy-width-sano}(b,d)), extending the monotonic relationship between width and bifurcation threshold into the plate-like regime. The Sano model captures a substantial fraction of the FEA shift at both widths, though it consistently underestimates the magnitude of the leftward migration. Interestingly, the $US \to S$ transition exhibits a qualitatively different behavior: the percentage shift saturates beyond $W/L = 1/6$, with the Sano model predicting marginal further reduction in critical shear values at $W/L = 1/3$ and $W/L = 1/2$ (see Figure~\ref{fig:homotopy-width-sano}(b-d), Table~\ref{tab:bifurcation_shifts_homotopy}). This contrasts with the FEA reference, where the $US \to S$ threshold continues to decrease significantly with increasing width. These observations are well quantified in Table~\ref{tab:bifurcation_shifts_homotopy}.

The existence of these branches confirms that the Sano energy landscape does support bifurcated equilibria at large widths. The difficulty encountered in earlier sections is not a fundamental limitation of the model but rather a consequence of the growing energy barrier at inflection points, which prevents the Newton-Raphson solver from escaping the basin of attraction surrounding the symmetric $U$ configuration during standard quasi-static loading. The homotopy-based approach overcomes this barrier by providing a continuous path in parameter space that connects the accessible bifurcated branch at narrow widths to the otherwise inaccessible branch at large widths, without requiring an external initial guess from a separate rod or plate simulation.

\begin{table}[t]
\centering
\small
\renewcommand{\arraystretch}{1.3}
\caption{Negative Percentage shift $-\Delta_\%$ in critical shear from their respective $W/L = 1/20$ baselines and absolute error $\delta_e$ relative to FEA, for the $U \to US$ and $US \to S$ transitions at $L/b = 100$.}
\label{tab:bifurcation_shifts_homotopy}
\begin{tabular}{@{} l l rr rr @{}}
\toprule
& & \multicolumn{2}{c}{$W/L = 1/3$} & \multicolumn{2}{c}{$W/L = 1/2$} \\
\cmidrule(lr){3-4} \cmidrule(lr){5-6}
Transition & Model & $-\Delta_\%$ (\%) & $\delta_e$ & $-\Delta_\%$ (\%) & $\delta_e$ \\
\midrule
\multirow{4}{*}{$U \to US$}
 & FEA (GT)    & 63.0 & $*$   & 72.2 & $*$   \\
 & Kirchhoff & 0.0  & 0.18 & 0.0  & 0.195 \\
 & Sano     & \textbf{35.7} & \textbf{0.08} & \textbf{46.4} & \textbf{0.075} \\
\midrule
\multirow{4}{*}{$US \to S$}
 & FEA (GT)     & 34.2 & $*$   & 47.6 & $*$   \\
 & Kirchhoff & 0.0  & 0.13 &  0.0  & 0.15 \\
 & Sano     & \textbf{18.4}  & \textbf{0.06} & \textbf{19.7} & \textbf{0.105} \\
\bottomrule
\end{tabular}
\end{table}

This loading strategy has broader implications beyond the specific benchmark studied here. In ribbon mechanics problems where multiple stable equilibria coexist but are separated by energy barriers that grow with geometric parameters, direct quasi-static continuation may fail to explore the full solution landscape. The homotopy approach of varying a geometric or material parameter along a path that connects an accessible branch to an inaccessible one, provides a systematic and solver-independent technique for discovering hidden equilibrium branches. Unlike seeding with an external initial guess, which requires a priori knowledge of the target configuration, the homotopy method requires only the identification of a parameter regime in which the bifurcation is accessible, making it applicable to problems where the target solution is not known in advance.

\section{Computational Cost Analysis}\label{sec:cost}
 
One-dimensional ribbon models offer a fundamental computational advantage over two-dimensional shell or plate discretizations.
A shell finite element model of a ribbon requires meshing both the length and the width, yielding a node count that scales as $\mathcal{O}(N_L \times N_W)$ and producing a sparse stiffness matrix whose bandwidth grows with $N_W$. By contrast, the discrete elastic ribbon models presented in this work discretize only the centerline, yielding $\mathcal{O}(N)$ degrees of freedom with a banded stiffness matrix of fixed bandwidth (11 DOFs per element stencil, except Wunderlich), and consequently $\mathcal{O}(N)$ assembly and solve cost per Newton-Raphson iteration. This dimensional reduction, from a two-dimensional mesh to a one-dimensional chain, constitutes the primary computational saving and is shared by all centerline-based models, including the standard Kirchhoff-based Discrete Elastic Rods (DER).
 
Within the family of one-dimensional models, the computational cost of the generalized Discrete Elastic Ribbon framework relative to standard DER depends on two factors: (i) the complexity of the constitutive energy function, and (ii) the difficulty of the nonlinear solve. The first factor arises from the chain-rule assembly in Algorithm~\ref{alg:general_energy}: Step~2 evaluates the energy $\mathcal{E}_k$, its gradient $\partial\mathcal{E}_k/\partial\boldsymbol{\epsilon}_k$, and its Hessian $\nabla^2_{\boldsymbol{\epsilon}_k}\mathcal{E}_k$ in strain space. For the Kirchhoff model, the energy is a quadratic function of individual strain components, so these derivatives are trivial and the constitutive Hessian is diagonal. For the coupled ribbon models (Sadowsky, Sano, Audoly), the cross-derivative $\partial^2\mathcal{E}_k/\partial\kappa^{(2)}\partial\tau$ is nonzero and the constitutive Hessian is dense in the $4 \times 4$ strain space, requiring additional floating-point operations per element. The Wunderlich model introduces further cost through the non-local finite-difference stencil for $\eta'$, which extends the element connectivity and increases the number of energy evaluations per assembly pass. The second factor is the adaptive time-stepping scheme (as mentioned in Algorithm~\ref{alg:implicit_euler}), which automatically reduces the time step near bifurcation points and instabilities to maintain convergence. Since ribbon models with stronger bending-twisting coupling tend to produce stiffer energy landscapes near inflection points, the solver may require more Newton-Raphson iterations and smaller time steps in precisely those regions where the physical behavior is most interesting. This adaptive refinement is essential for robustness but increases the total step count relative to a simulation that encounters no instabilities.
 
These two factors, constitutive complexity and solver difficulty, affect wall-clock time but not the asymptotic scaling with mesh size, which remains $\mathcal{O}(N)$ for all models. To isolate and quantify their practical impact, we developed a high-performance reimplementation of the framework in JAX~\cite{jax2018github}, which preserves the same algorithmic structure (implicit Euler integration, Newton-Raphson iteration with adaptive time-stepping, and the robust linear solver of Algorithm~\ref{alg:robust_solve}) while eliminating the overhead of the PyDiSMech-based~\cite{lahoti2025py} NumPy-PyTorch framework. The key advantages of the JAX implementation are: (i) vectorized per-element assembly via \texttt{jax.vmap}, which fuses the constitutive and geometric derivative stages into a single compiled kernel; (ii) JIT compilation of the entire Newton-Raphson inner loop, eliminating Python interpreter overhead; and (iii) a unified tracing framework for both analytical geometric derivatives and autodiff-based constitutive derivatives, avoiding cross-framework data transfers. Full implementation details are provided in~\ref{app:jax_implementation}.
 
\subsection{Efficiency Benchmarks}
 
We benchmark the JAX implementation on the shear-induced bifurcation problem of Section~\ref{sec:results} at $L/b = 100$, focusing on the shear phase (corresponding to $4.95$\,s of simulated time) during which the bifurcation and post-bifurcation dynamics occur. We compare three energy models---Kirchhoff, Sano, and Audoly---across four ribbon widths $W/L \in \{1/40, 1/20, 1/12, 1/6\}$ and two mesh resolutions $N = 45$ and $N = 63$ nodes along centerline length $L=100 \, \text{mm}$. All simulations were performed on a single core of an AMD Ryzen Threadripper PRO 7975WX (32 cores, 3.5\,GHz base / 5.3\,GHz boost). The results are summarized in Tables~\ref{tab:efficiency_N45} and~\ref{tab:efficiency_N63}.
 
\begin{table}[htbp]
\centering
\small
\renewcommand{\arraystretch}{1.2}
\caption{Efficiency benchmarks for the shear-induced bifurcation problem at $N = 45$ nodes, $L = 100 \, \textrm{mm}$ and $L/b = 100$ for ribbon widths $W/L \in \{1/40, 1/20, 1/12, 1/6\}$. Columns report: real-time ratio ($\times$Sim = wall-clock time\,/\,simulated time; lower is faster), wall-clock time in seconds, total time steps (or Steps), total Newton-Raphson iterations (or \#Iterations), and average Newton-Raphson iterations per step (or Iterations/step). }
\label{tab:efficiency_N45}
\begin{tabular}{@{} ll rrrrr @{}}
\toprule
Model & $W/L$ & $\times$Sim & Wall-Clock\,(s) & Steps & \#Iterations & Iterations/step \\
\midrule
\multirow{4}{*}{Kirchhoff}
 & 1/40 & 0.32 &  1.58 &  285 & 1456 & 5.1 \\
 & 1/20 & 0.52 &  2.57 &  604 & 3426 & 5.7 \\
 & 1/12 & 0.59 &  2.93 &  513 & 3140 & 6.1 \\
 & 1/6  & 2.16 & 10.71 & 1436 & 8921 & 6.2 \\
\midrule
\multirow{4}{*}{Sano}
 & 1/40 & 0.31 &  1.51 &  284 & 1448 & 5.1 \\
 & 1/20 & 0.61 &  3.00 &  708 & 3831 & 5.4 \\
 & 1/12 & 0.58 &  2.89 &  497 & 2963 & 6.0 \\
 & 1/6  & 1.65 &  8.15 & 1164 & 7512 & 6.5 \\
\midrule
\multirow{4}{*}{Audoly}
 & 1/40 & 0.26 &  1.31 &  214 & 1140 & 5.3 \\
 & 1/20 & 1.57 &  7.77 & 1516 &10346 & 6.8 \\
 & 1/12 & 0.72 &  3.55 &  471 & 2906 & 6.2 \\
 & 1/6  & 3.39 & 16.77 & 1430 & 8496 & 5.9 \\
\bottomrule
\end{tabular}
\end{table}
 
\begin{table}[htbp]
\centering
\small
\renewcommand{\arraystretch}{1.2}
\caption{Efficiency benchmarks for the shear-induced bifurcation problem at $N = 63$ nodes $L = 100 \, \textrm{mm}$ and $L/b = 100$ for ribbon widths $W/L \in \{1/40, 1/20, 1/12, 1/6\}$. Column definitions are the same as in Table~\ref{tab:efficiency_N45}.}
\label{tab:efficiency_N63}
\begin{tabular}{@{} ll rrrrr @{}}
\toprule
Model & $W/L$ & $\times$Sim & Wall-Clock\,(s) & Steps & \#Iterations & Iterations/step \\
\midrule
\multirow{4}{*}{Kirchhoff}
 & 1/40 & 0.35 &  1.75 &  338 & 1832 & 5.4 \\
 & 1/20 & 0.72 &  3.55 &  786 & 4653 & 5.9 \\
 & 1/12 & 1.83 &  9.04 & 1208 & 7924 & 6.6 \\
 & 1/6  & 8.23 & 40.76 & 2535 &18487 & 7.3 \\
\midrule
\multirow{4}{*}{Sano}
 & 1/40 & 0.32 &  1.59 &  306 & 1673 & 5.5 \\
 & 1/20 & 0.63 &  3.12 &  630 & 3864 & 6.1 \\
 & 1/12 & 2.01 &  9.96 & 1118 & 7097 & 6.3 \\
 & 1/6  & 6.91 & 34.22 & 2247 &15758 & 7.0 \\
\midrule
\multirow{4}{*}{Audoly}
 & 1/40 & 0.30 &  1.46 &  265 & 1454 & 5.5 \\
 & 1/20 & 1.26 &  6.22 &  763 & 4569 & 6.0 \\
 & 1/12 & 2.80 & 13.87 & 2611 &16898 & 6.5 \\
 & 1/6  & 8.29 & 41.05 & 2181 &14506 & 6.7 \\
\bottomrule
\end{tabular}
\end{table}
 
\subsection{Discussion}
 
Several observations emerge from the benchmark results. At narrow widths ($W/L = 1/40$), all three models run roughly three times faster than real time ($\times\text{Sim} \leq 0.35$) at both mesh resolutions, with wall-clock times of 1.3--1.8\,s. This confirms that the one-dimensional framework is computationally inexpensive in regimes where the nonlinear solve is well-conditioned.
 
The Kirchhoff, Sano, and Audoly models exhibit comparable per-iteration costs at a given mesh resolution. At $N = 45$ and $W/L = 1/6$, for example, Kirchhoff and Sano yield per-step wall-clock times of ${\sim}7.5$\,ms and ${\sim}7.0$\,ms respectively, or equivalently ${\sim}1.2$\,ms and ${\sim}1.1$\,ms per Newton iteration. The cost of the coupled constitutive Hessian in the Sano and Audoly models is mostly negligible compared with the geometric assembly and the banded linear solve. This is expected since the constitutive Hessian is a $4 \times 4$ dense matrix evaluated per element, whereas the dominant per-step cost lies in the $11 \times 11$ local-to-global scatter into banded storage and the banded factorization itself.

The total computational cost is dominated by the adaptive time-stepping response rather than per-step expense. Wider ribbons produce stiffer energy landscapes near bifurcation points, triggering smaller time steps and additional Newton iterations per step; this is the mechanism behind the sharp growth in wall-clock time from $W/L = 1/20$ to $W/L = 1/6$ at both mesh sizes. Within this regime the ordering of models is driven primarily by their step count rather than their per-step cost. At $W/L = 1/6$, the Sano model is the most efficient of the three at both mesh resolutions (8.2\,s at $N=45$, 34.2\,s at $N=63$), illustrating that the bending-twisting coupling term can improve not only accuracy but also computational efficiency relative to the Kirchhoff baseline near instabilities, most probably because the coupled energy provides a smoother path through the bifurcation and reduces the number of failed Newton iterations and time-step reductions. The Audoly model is consistently the most expensive at multiple widths, most probably because of its constitutive complexity coming from the transition function $\varphi(v)$ which introduces curvature-dependent stiffness where the model interpolates between its Sano and Sadowsky limits, which the adaptive time-stepper resolves by taking smaller time steps or higher per-iteration cost.

Increasing the mesh resolution from $N = 45$ to $N = 63$ (a $1.4\times$ increase in DOFs) produces wall-clock ratios that vary with width: $1.05$--$1.11\times$ at $W/L = 1/40$ and up to $3$--$4\times$ at $W/L = 1/6$. To isolate the linear-algebra cost from step-count effects, we normalize by the total number of Newton iterations: the median wall clock time per NR iteration ratio across all twelve model-width configurations is approximately $1.1\times$, at or below the ideal $1.4\times$ linear-in-DOF bound. The residual super-linear wall-time growth at wider widths is therefore almost entirely attributable to the adaptive time-stepper taking more steps at the finer mesh, reflecting the spatial resolution of stiff bifurcation features rather than any growth in the per-iteration solve cost. This confirms that the banded factorization recovers the asymptotic $\mathcal{O}(N)$ per-Newton-iteration cost predicted by the one-dimensional discretization with fixed bandwidth.
 
In summary, the generalized Discrete Elastic Ribbon framework introduces negligible per-iteration overhead relative to standard DER for the coupled energy models (Kirchhoff, Sano, and Audoly per-Newton-iteration costs differ by less than 15\%), and the banded factorization recovers the expected $\mathcal{O}(N)$ per-iteration scaling. The dominant cost variation across geometries arises from the adaptive time-stepping response to the stiffness of the nonlinear problem near bifurcation. The JAX implementation achieves faster than real-time performance for narrow ribbons at both mesh resolutions and remains tractable even for the widest ribbon tested ($N = 63$, $W/L = 1/6$) completing under 42\,s of wall-clock time on a single CPU core.

\section{Relevant Benchmarks}\label{sec:relevant_benchmarks}


The shear-induced bifurcation study of Section~\ref{sec:results} established the Sano model as the most accurate one-dimensional formulation at narrow to moderate widths, while exposing the inability of the strict developable models (Sadowsky and Wunderlich) to spontaneously traverse the pitchfork. To test whether these conclusions generalize beyond a single loading scenario, we consider two additional boundary-value problems within the same clamped pre-buckled geometry: (i) a pure boundary twist applied to the pre-buckled ribbon, and (ii) a combined shear and twist loading in which both kinematic inputs are imposed simultaneously. Because the Wunderlich model fails to converge reliably under these more aggressive loadings, we restrict the comparison to four one-dimensional models---Kirchhoff, Sadowsky, Sano, and Audoly---benchmarked against shell-element FEA.

\subsection{Twist Induced Response}

\begin{figure}[htbp]
    \centering
    \includegraphics[width=1.0\textwidth]{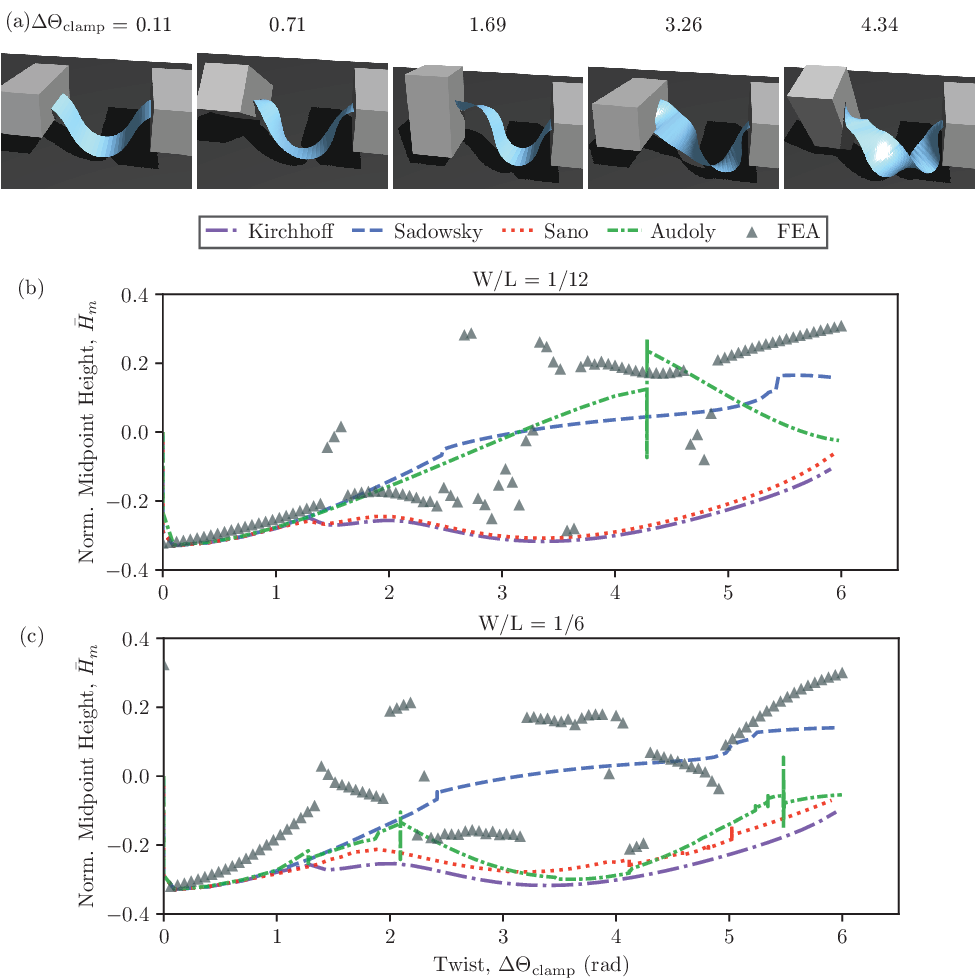}
    \caption{Twist-induced response of the pre-buckled ribbon. (a) Renderings of representative deformed configurations at $W/L = 1/6$ using Sano energy. Normalized midpoint height $\bar{H}_m$ versus clamp twist $\Delta\Theta_{\text{clamp}}$ for (b) $W/L = 1/12$ and (c) $W/L = 1/6$, comparing four one-dimensional models against FEA at $L/b = 100$.}    \label{fig:twist-bc}
\end{figure}

The pre-buckling stage follows the same clamped longitudinal compression protocol described in Section~\ref{sec:results}, yielding the symmetric $U$-shaped equilibrium at $\Delta L/L = 1/4$. Holding this compressed span fixed, as shown in Figure~\ref{fig:twist-bc}a, we then rotate one clamped end about the longitudinal axis by a clamp twist angle, $\Delta\Theta_{\text{clamp}}$. As before, we track the normalized midpoint height $\bar{H}_m = H_m/L$ (here retaining the sign, unlike the absolute value used in Section~\ref{sec:results}) and compare the four one-dimensional models against FEA at $L/b = 100$ for two widths, $W/L = 1/12$ and $W/L = 1/6$. The material parameters ($Y = 10$~GPa, $\nu = 0.5$) are identical to those of Section~\ref{sec:results}.

The FEA twist--displacement response in Figure~\ref{fig:twist-bc} exhibits three regimes. First, $\bar{H}_m$ rises monotonically with $\Delta\Theta_{\text{clamp}}$ as the pre-buckled arch stiffens under twist. At a critical twist the response splits into an upper and a lower branch; this branching is visibly more pronounced at $W/L = 1/6$ than at $W/L = 1/12$. As the twist increases further, the two branches merge and $\bar{H}_m$ resumes a monotonic climb on the merged path.
 
At a coarse level, none of the one-dimensional models quantitatively reproduce the FEA twist--displacement curves at either width. At $W/L = 1/12$ (Figure~\ref{fig:twist-bc}b), all four models remain close to one another and to FEA up to $\Delta\Theta_{\text{clamp}} \approx 1.4 \, \text{rad}$. Beyond this point the models diverge: Sadowsky traces a path that lies between the FEA upper and lower branches, Kirchhoff and Sano track closer to the lower branch with the discrepancy growing monotonically. Audoly follows Sadowsky until $\Delta\Theta_{\text{clamp}} \approx 3.2 \, \text{rad}$ before departing into a qualitatively different trajectory. At $W/L = 1/6$ (Figure~\ref{fig:twist-bc}c), Audoly follows Sadowsky until $\Delta\Theta_{\text{clamp}} \approx 2.15 \, \text{rad}$, and then aligns with Sano for the remainder of the loading, reinforcing the interpolant character between the Sadowsky and Sano limits already noted in Section~\ref{sec:results}.

Despite the quantitative discrepancies, the qualitative structure of the FEA response, characterized by an initial climb, a partial drop along the lower branch, and a subsequent increase, is reproduced by Kirchhoff and Sano at $W/L = 1/12$, and by Kirchhoff, Sano, and Audoly at $W/L = 1/6$. The FEA width effect, whereby a wider ribbon attains a larger $\bar{H}_m$ at the same $\Delta\Theta_{\text{clamp}}$, is captured by Sano but absent in Kirchhoff, again depicting its width-invariant characteristic. A subtler width effect appears in the early-twist regime ($\Delta\Theta_{\text{clamp}} < 1.2 \, \text{rad}$): at $W/L = 1/12$ the initial slope $\frac{\partial\bar{H}_m}{\partial\Delta\Theta_{\text{clamp}}}$ predicted by all four one-dimensional models agrees with FEA, but at $W/L = 1/6$ the FEA slope is noticeably steeper than any one-dimensional prediction. This discrepancy suggests that the absolute-stiffness contribution of width, distinct from the kinematic shift of bifurcation thresholds, is not faithfully encoded in any of the centerline-based formulations tested.

In summary, Sano and Kirchhoff provide the closest match to FEA at $W/L = 1/12$, while Audoly is nearest at $W/L = 1/6$. Nevertheless, even for the closest match a significant quantitative gap persists at both widths.

\subsection{Shear-plus-Twist Induced Response}

\begin{figure}[htbp]
    \centering
    \includegraphics[width=1.0\textwidth]{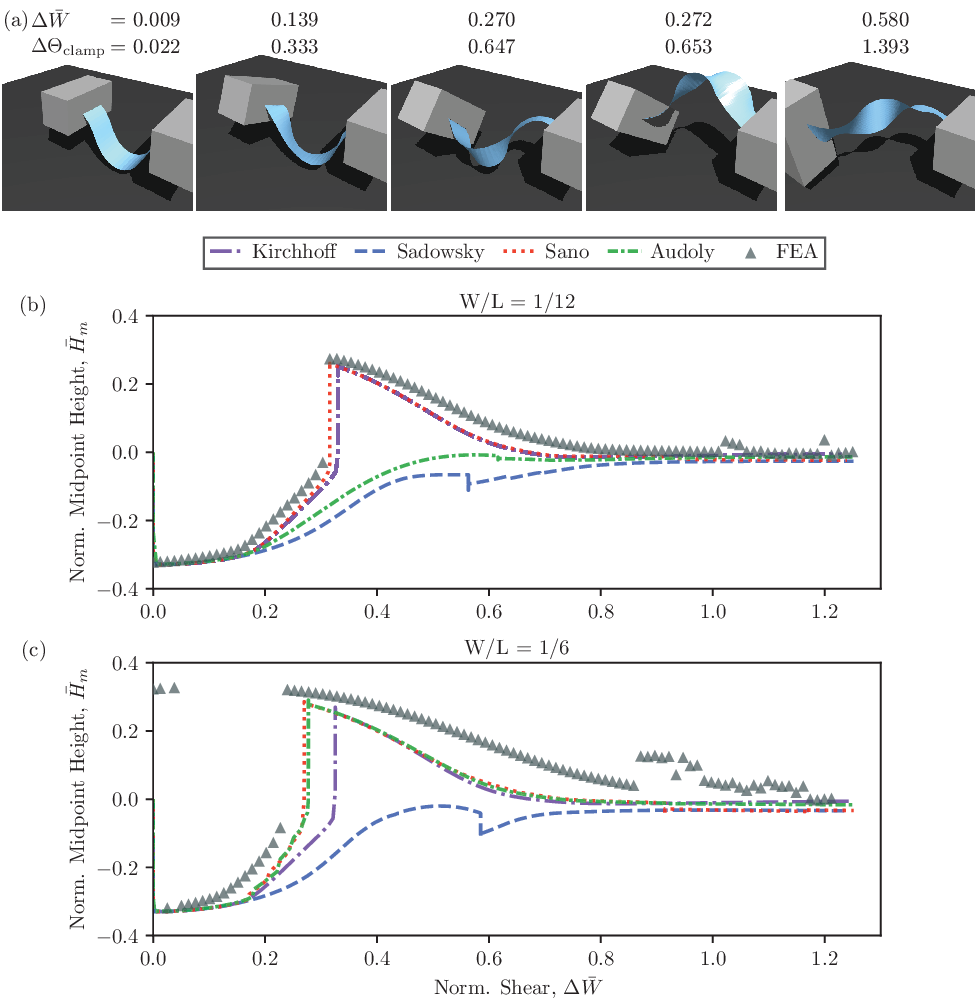}
    \caption{Combined shear and twist loading of the pre-buckled ribbon. (a) Renderings of representative deformed configurations at $W/L = 1/6$ using Sano energy. Normalized midpoint height $\bar{H}_m$ versus normalized transverse shear $\Delta\bar{W}$ for (b) $W/L = 1/12$ and (c) $W/L = 1/6$, comparing four one-dimensional models against FEA at $L/b = 100$.}    \label{fig:shear-plus-twist-bc}
\end{figure}

The third benchmark combines the two kinematic inputs studied above. Starting from the compressed $U$-shape, we simultaneously impose a transverse shear $\Delta\bar{W}$ and a boundary twist $\Delta\Theta_{\text{clamp}}$ at one clamped end, with both increments ramped concurrently at a fixed ratio (Figure~\ref{fig:shear-plus-twist-bc}a). We again compare Kirchhoff, Sadowsky, Sano, and Audoly against FEA at $L/b = 100$ for $W/L \in \{1/12,\, 1/6\}$, tracking $\bar{H}_m$ as the combined loading progresses.

The FEA response in Figure~\ref{fig:shear-plus-twist-bc}(b,c) exhibits a characteristic three-stage profile. First, $\bar{H}_m$ rises monotonically as the combined shear and twist accumulate, stiffening the pre-buckled arch. At a critical load the arch snaps abruptly to the opposite side, after which $\bar{H}_m$ decays and asymptotes toward zero. Three features serve as diagnostics for model comparison: the pre-snap slope $\frac{\partial\bar{H}_m}{\partial\Delta\bar{W}}$, the snap location in $\Delta\bar{W}$, and the post-snap decay rate.
 
At $W/L = 1/12$ (Figure~\ref{fig:shear-plus-twist-bc}b), both Kirchhoff and Sano track the FEA profile closely, with Sano providing the most accurate match in both the pre-snap slope and the snap location; minor deviations appear only in the post-snap decay. Sadowsky fails outright, missing the snap entirely. Audoly again acts as an interpolant between Sadowsky and Sano, sitting closer to Sadowsky at this width. At $W/L = 1/6$ (Figure~\ref{fig:shear-plus-twist-bc}c), the FEA profile shifts leftward, consistent with the width-induced reduction of the critical load observed throughout this work. Sano reproduces this leftward shift, though a residual gap remains in both the initial slope and the snap location. Kirchhoff, as in all previous benchmarks, displays no width sensitivity and consequently underestimates the shift. Audoly at this wider geometry coincides with Sano, consistent with its interpolant character migrating toward Sano as the width increases.

\subsection{Summary of Additional Benchmarks}
 
The combined results from the twist and shear-plus-twist benchmarks corroborate the ranking established by the shear-induced bifurcation study of Section~\ref{sec:results}. Across all three boundary-value problems, Sano emerges as the best-performing one-dimensional model: it captures both the qualitative instability behavior and the width-dependent shift of critical thresholds. Sadowsky consistently fails to traverse instabilities involving inflection points, Kirchhoff is width-invariant, and Audoly interpolates between the Sadowsky and Sano limits with a width-dependent bias that shifts progressively toward Sano as the ribbon widens. This concordance across three distinct boundary value problems strengthens the conclusion that Sano is the most reliable one-dimensional baseline for pre-buckled ribbon mechanics, while reaffirming that a quantitative gap relative to shell FEA persists and widens as the ribbon enters the plate-like regime.

\section{Conclusion}\label{sec:conclusion}

This work introduces a generalized discrete differential geometry framework in which the elastic potential is expressed as an arbitrary function of the coupled strain vector $\boldsymbol{\epsilon} = [\varepsilon, \kappa^{(1)}, \kappa^{(2)}, \tau]^T$. By deriving exact gradients and Hessians that retain all cross-derivative contributions $\partial^2 E / \partial \epsilon_j \partial \epsilon_k$, the formulation accommodates any one-dimensional constitutive law with nonlinearly coupled strains. The clean separation between model-independent discrete geometry and model-specific scalar energy $\mathcal{E}(\boldsymbol{\epsilon})$ ensures that incorporating a new energy model requires only specifying the energy as a function of strains, with no modification to the geometric framework.
 
Simulating the singular energy landscapes of ribbon theories demands more than standard Newton-Raphson iteration. Our implicit Euler scheme regularizes equilibrium through inertial terms that enable traversal of energy barriers near bifurcation points. Adaptive time-stepping concentrates resolution near snap-through events, and Tikhonov-Miller regularization ensures positive definiteness when the Hessian becomes near-singular. Together, these techniques yield robust convergence across all five energy models, including formulations that have historically required specialized continuation methods.
 
Within this unified environment, we discretized five prominent ribbon energy formulations: Kirchhoff, Sadowsky, Wunderlich, Sano, \& Audoly, and validated them against shell-element FEA on three benchmark problems: shear-induced pitchfork bifurcation, boundary twist, and combined shear-plus-twist loading of pre-buckled bands. The strict developable models (Sadowsky and Wunderlich) remain trapped on the symmetric branch under all conditions tested, confirming that the inflection-point energy barrier is intrinsic to the developability constraint. Their use should therefore be restricted to configurations known a priori to remain free of inflection points with nonzero twist, or to post-processing with externally seeded initial guesses. The Kirchhoff model captures bifurcations and snap-through events but is invariant to width changes, making it a reliable choice only for narrow ribbons ($W/L \lesssim 1/20$)~\cite{yu2019bifurcations, huang2020shear}, where it matches the accuracy of the more complex ribbon models while offering a simpler decoupled quadratic energy structure. Among all formulations, the Sano model achieves the highest fidelity across the full suite of benchmarks, consistently capturing width-dependent shifts of critical thresholds, including the leftward migration of bifurcation points under shear, the width-sensitive twist--displacement response, and the leftward shift of snap locations under combined loading, with the lowest absolute errors relative to FEA across $W/L \in \{1/20, 1/12, 1/6\}$. The Audoly model offers comparable accuracy to Sano at specific width-thickness combinations but exhibits inconsistent performance across the parameter space, particularly due its interpolant behavior between Sano and Sadowsky. Beyond $W/L \approx 1/3$, none of the tested models can spontaneously access bifurcated branches under direct quasi-static loading; however, our homotopy-based width transition successfully uncovered hidden bifurcation branches at $W/L = 1/3$ and $1/2$ within the Sano energy landscape, confirming that the equilibrium structure persists but is shielded by growing energy barriers. The gap between even the best one-dimensional prediction and FEA widens as the ribbon enters the plate-like regime, exposing a fundamental ceiling on centerline-based models.

Efficiency benchmarks using a high-performance JAX reimplementation confirm that the coupled constitutive Hessian in the Sano model introduces less than 15\% per-Newton-iteration overhead relative to the Kirchhoff baseline. At wider widths where bifurcation occurs, the Sano model is the most efficient of the three models tested, as the bending-twisting coupling provides a smoother energy landscape that reduces the number of failed Newton iterations and adaptive time-step reductions. The Audoly model, by contrast, incurs consistently higher computational cost at multiple widths due to the additional stiffness that its transition function $\varphi(v)$ introduces at intermediate curvatures. Taken together, the accuracy and efficiency results establish the Sano model as the recommended default one-dimensional formulation for ribbons in the range $1/20 \lesssim W/L \lesssim 1/6$. At narrower widths ($W/L \lesssim 1/20$), both Kirchhoff and Sano achieve equivalent accuracy and efficiency, and Kirchhoff's simpler energy structure makes it the easier model to implement.
 
These findings open several directions for future work. The quantitative shortfall of the Sano model at large widths motivates the development of enhanced one-dimensional formulations that encode higher-order width-dependent mechanics, potentially through correction terms calibrated against the Sano baseline. The modular architecture of the framework, where the constitutive response enters solely through $\mathcal{E}(\boldsymbol{\epsilon})$ and its derivatives, is naturally suited to data-driven approaches: a neural network trained on high-fidelity shell data could replace the analytical energy function while inheriting the exact gradient and Hessian propagation, yielding a learned one-dimensional model that operates within the established Discrete Elastic Rod geometric framework. Further validation on complementary boundary conditions, including twist-induced snapping~\cite{sano2019twist, huang2022discrete}, bifurcation under twist~\cite{feng2025twist}, and snap-through instabilities in curved ribbons, would continue to map the domain of validity of each energy model and guide formulation selection for specific engineering applications.

\section{Acknowledgments}
We acknowledge financial support from the National Science Foundation (grant numbers: 2047663 and 2209782). We thank Radha Lahoti for helping with the DiSMech software environment. We also thank Ryan Chaiyakul for initiating the efforts towards the JAX implementation.

\section{Data Availability}
The source code for our framework is publicly available at \url{https://github.com/StructuresComp/discrete-elastic-ribbon} and \url{https://github.com/StructuresComp/discrete-elastic-ribbon-jax}.

\section{Declaration of generative AI and AI-assisted technologies in the manuscript preparation process}

During the preparation of this work the author(s) used ChatGPT (OpenAI) for assistance with language editing and improving clarity. After using this tool/service, the author(s) reviewed and edited the content as needed and take(s) full responsibility for the content of the published article.

\appendix

\section{Derivation of Strain Gradients}
\label{app:strain_gradients}

This appendix provides the analytical expressions for the geometric gradients $\mathbf{G}_k = \partial\boldsymbol{\epsilon}_k/\partial\mathbf{q}[\mathcal{I}_k]$ and Hessians $\mathbf{H}_{\epsilon,k} = \partial^2\boldsymbol{\epsilon}_k/\partial\mathbf{q}[\mathcal{I}_k]^2$ of each strain measure at element $k$ with respect to its 11 local DOFs. These are model-independent geometric quantities.

\subsection{Gradient of Axial Strain}

The axial strain on edge $i$ is $\varepsilon^i = \|\mathbf{e}^i\| / \|\overline{\mathbf{e}}^i\| - 1$. The gradient with respect to the nodal positions is,
\begin{equation}
\frac{\partial \varepsilon^i}{\partial \mathbf{x}_i} = -\frac{\mathbf{t}^i}{\|\overline{\mathbf{e}}^i\|}, \qquad \frac{\partial \varepsilon^i}{\partial \mathbf{x}_{i+1}} = \frac{\mathbf{t}^i}{\|\overline{\mathbf{e}}^i\|},
\end{equation}
where $\mathbf{t}^i = \mathbf{e}^i / \|\mathbf{e}^i\|$ is the unit tangent.

\subsection{Gradient of Curvature}

The discrete curvature components $\kappa_i^{(1)}$ and $\kappa_i^{(2)}$ depend on three consecutive nodes $\{\mathbf{x}_{i-1}, \mathbf{x}_i, \mathbf{x}_{i+1}\}$ and two twist angles $\{\theta^{i-1}, \theta^i\}$. The gradients are
\begin{equation}
\frac{\partial \kappa_i^{(\alpha)}}{\partial \mathbf{x}_j} = \frac{\partial \kappa_i^{(\alpha)}}{\partial (\boldsymbol{\kappa}_b)_i} \cdot \frac{\partial (\boldsymbol{\kappa}_b)_i}{\partial \mathbf{x}_j} + \frac{\partial \kappa_i^{(\alpha)}}{\partial \mathbf{m}_\beta^{i-1}} \cdot \frac{\partial \mathbf{m}_\beta^{i-1}}{\partial \mathbf{x}_j} + \frac{\partial \kappa_i^{(\alpha)}}{\partial \mathbf{m}_\beta^{i}} \cdot \frac{\partial \mathbf{m}_\beta^{i}}{\partial \mathbf{x}_j} \ ,
\end{equation}
for $j \in \{i-1, i, i+1\}$ and $\alpha, \beta \in \{1, 2\}$.

The gradient of the curvature binormal with respect to nodal positions follows from differentiation of the cross product formula. Detailed expressions are provided in \cite{jawed2018primer}.

\subsection{Gradient of Twist}

The discrete twist $\tau_i = \theta^i - \theta^{i-1} + m_{\text{ref}}^i$ has a direct dependence on the twist angles
\begin{equation}
\frac{\partial \tau_i}{\partial \theta^{i-1}} = -1, \qquad \frac{\partial \tau_i}{\partial \theta^i} = 1.
\end{equation}

The reference twist $m_{\text{ref}}^i$ depends on the nodal positions through the parallel transport of the reference frame, yielding additional terms
\begin{equation}
\frac{\partial \tau_i}{\partial \mathbf{x}_j} = \frac{\partial m_{\text{ref}}^i}{\partial \mathbf{x}_j}, \quad j \in \{i-1, i, i+1\}.
\end{equation}

\section{Derivation of Energy Hessian}
\label{app:hessian_derivation}

This appendix derives the local Hessian $\nabla^2_{\mathbf{q}[\mathcal{I}_k]} \mathcal{E}_k$ 
for a single element $k$, corresponding to the constitutive and geometric stiffness terms assembled in Equation~\eqref{eq:localHessian}.

\subsection{Expansion via Product Rule}

Starting from the local gradient of element $k$
\begin{equation}
\nabla_{\mathbf{q}[\mathcal{I}_k]} \mathcal{E}_k = 
\sum_{l=1}^{4} \frac{\partial \mathcal{E}_k}{\partial \epsilon_{k,l}} 
\mathbf{G}_{k,l} \ ,
\end{equation}
where $\mathbf{G}_{k,l}$ denotes the $l$-th row of $\mathbf{G}_k$, 
i.e., the gradient of the $l$-th strain component with respect to 
$\mathbf{q}[\mathcal{I}_k]$.

\subsection{Chain Rule for Energy Derivatives}

The gradient of each partial derivative of the element energy 
$\mathcal{E}_k$ with respect to the local DOFs is computed via 
the chain rule
\begin{equation}
\nabla_{\mathbf{q}[\mathcal{I}_k]}\!\left(
\frac{\partial \mathcal{E}_k}{\partial \epsilon_{k,l}}\right) = 
\sum_{m=1}^{4} \frac{\partial^2 \mathcal{E}_k}
{\partial \epsilon_{k,l}\, \partial \epsilon_{k,m}} \mathbf{G}_{k,m} \ ,
\end{equation}
where $\mathbf{G}_{k,m}$ denotes the gradient of the $m$-th strain 
component of element $k$ with respect to $\mathbf{q}[\mathcal{I}_k]$.

\subsection{Complete Hessian Expression}

Substituting and reorganizing yields the local Hessian of element $k$,
which corresponds to the two terms in Equation~\eqref{eq:localHessian}
\begin{align}
\nabla^2_{\mathbf{q}[\mathcal{I}_k]} \mathcal{E}_k 
&= \sum_{l=1}^{4} \frac{\partial^2 \mathcal{E}_k}{\partial \epsilon_{k,l}^2} 
\mathbf{G}_{k,l} \otimes \mathbf{G}_{k,l} 
+ \sum_{l=1}^{4} \frac{\partial \mathcal{E}_k}{\partial \epsilon_{k,l}} 
\mathbf{H}_{\epsilon_{k,l}} \nonumber \\
&+ \sum_{l < m} \frac{\partial^2 \mathcal{E}_k}
{\partial \epsilon_{k,l}\, \partial \epsilon_{k,m}} 
\left( \mathbf{G}_{k,l} \otimes \mathbf{G}_{k,m} + 
       \mathbf{G}_{k,m} \otimes \mathbf{G}_{k,l} \right) \ ,
\end{align}
where the first line corresponds to the diagonal constitutive and 
geometric stiffness terms, and the second line to the off-diagonal 
constitutive coupling between strain components. The global Hessian 
$\mathbf{H}_E$ is then assembled via Equation~\eqref{eq:localHessian} 
by summing over all elements $k$.

Expanding explicitly for $\boldsymbol{\epsilon}_k = 
[\varepsilon_k,\, \kappa^{(1)}_k,\, \kappa^{(2)}_k,\, \tau_k]^T$:

\textbf{Diagonal terms:}
\begin{align}
&\frac{\partial^2 \mathcal{E}_k}{\partial \varepsilon_k^2} 
 \mathbf{G}_{k,\varepsilon} \otimes \mathbf{G}_{k,\varepsilon} + 
 \frac{\partial \mathcal{E}_k}{\partial \varepsilon_k} 
 \mathbf{H}_{k,\varepsilon} \\
&\frac{\partial^2 \mathcal{E}_k}{\partial (\kappa^{(1)}_k)^2} 
 \mathbf{G}_{k,\kappa^{(1)}} \otimes \mathbf{G}_{k,\kappa^{(1)}} + 
 \frac{\partial \mathcal{E}_k}{\partial \kappa^{(1)}_k} 
 \mathbf{H}_{k,\kappa^{(1)}} \\
&\frac{\partial^2 \mathcal{E}_k}{\partial (\kappa^{(2)}_k)^2} 
 \mathbf{G}_{k,\kappa^{(2)}} \otimes \mathbf{G}_{k,\kappa^{(2)}} + 
 \frac{\partial \mathcal{E}_k}{\partial \kappa^{(2)}_k} 
 \mathbf{H}_{k,\kappa^{(2)}} \\
&\frac{\partial^2 \mathcal{E}_k}{\partial \tau_k^2} 
 \mathbf{G}_{k,\tau} \otimes \mathbf{G}_{k,\tau} + 
 \frac{\partial \mathcal{E}_k}{\partial \tau_k} 
 \mathbf{H}_{k,\tau}.
\end{align}

\textbf{Cross terms:}
\begin{align}
&\frac{\partial^2 \mathcal{E}_k}{\partial \varepsilon_k \partial \kappa^{(1)}_k} 
 \left( \mathbf{G}_{k,\varepsilon} \otimes \mathbf{G}_{k,\kappa^{(1)}} + 
        \mathbf{G}_{k,\kappa^{(1)}} \otimes \mathbf{G}_{k,\varepsilon} \right) \\
&\frac{\partial^2 \mathcal{E}_k}{\partial \varepsilon_k \partial \kappa^{(2)}_k} 
 \left( \mathbf{G}_{k,\varepsilon} \otimes \mathbf{G}_{k,\kappa^{(2)}} + 
        \mathbf{G}_{k,\kappa^{(2)}} \otimes \mathbf{G}_{k,\varepsilon} \right) \\
&\frac{\partial^2 \mathcal{E}_k}{\partial \varepsilon_k \partial \tau_k} 
 \left( \mathbf{G}_{k,\varepsilon} \otimes \mathbf{G}_{k,\tau} + 
        \mathbf{G}_{k,\tau} \otimes \mathbf{G}_{k,\varepsilon} \right) \\
&\frac{\partial^2 \mathcal{E}_k}{\partial \kappa^{(1)}_k \partial \kappa^{(2)}_k} 
 \left( \mathbf{G}_{k,\kappa^{(1)}} \otimes \mathbf{G}_{k,\kappa^{(2)}} + 
        \mathbf{G}_{k,\kappa^{(2)}} \otimes \mathbf{G}_{k,\kappa^{(1)}} \right) \\
&\frac{\partial^2 \mathcal{E}_k}{\partial \kappa^{(1)}_k \partial \tau_k} 
 \left( \mathbf{G}_{k,\kappa^{(1)}} \otimes \mathbf{G}_{k,\tau} + 
        \mathbf{G}_{k,\tau} \otimes \mathbf{G}_{k,\kappa^{(1)}} \right) \\
&\frac{\partial^2 \mathcal{E}_k}{\partial \kappa^{(2)}_k \partial \tau_k} 
 \left( \mathbf{G}_{k,\kappa^{(2)}} \otimes \mathbf{G}_{k,\tau} + 
        \mathbf{G}_{k,\tau} \otimes \mathbf{G}_{k,\kappa^{(2)}} \right).
\end{align}

\subsection{Remarks on Structure}

For the Kirchhoff model, all cross terms vanish since 
$\partial^2 \mathcal{E}_k / \partial \epsilon_{k,l}\, \partial \epsilon_{k,m} = 0$ 
for $l \neq m$. The local Hessian reduces to
\begin{equation}
\nabla^2_{\mathbf{q}[\mathcal{I}_k]} \mathcal{E}_k^{\text{Kirc}} = 
\sum_{l=1}^{4} \frac{\partial^2 \mathcal{E}_k}{\partial \epsilon_{k,l}^2} 
\mathbf{G}_{k,l} \otimes \mathbf{G}_{k,l} + 
\sum_{l=1}^{4} \frac{\partial \mathcal{E}_k}{\partial \epsilon_{k,l}} 
\mathbf{H}_{k,l}.
\end{equation}
For ribbon models (Sadowsky, Sano, Audoly), the dominant coupling 
occurs between $\kappa^{(2)}_k$ and $\tau_k$, requiring evaluation of
\begin{equation}
\frac{\partial^2 \mathcal{E}_k}{\partial \kappa^{(2)}_k\, \partial \tau_k} 
\left( \mathbf{G}_{k,\kappa^{(2)}} \otimes \mathbf{G}_{k,\tau} + 
       \mathbf{G}_{k,\tau} \otimes \mathbf{G}_{k,\kappa^{(2)}} \right).
\end{equation}
The geometric terms $\mathbf{G}_{k,l}$ and $\mathbf{H}_{k,l}$ remain 
unchanged across all energy models, while the energy-dependent scalars 
$\partial \mathcal{E}_k / \partial \epsilon_{k,l}$ and 
$\partial^2 \mathcal{E}_k / \partial \epsilon_{k,l}\, \partial \epsilon_{k,m}$ 
are model-specific. The global Hessian $\mathbf{H}_E$ follows by 
assembling each local block via Equation~\eqref{eq:localHessian}.

\section{JAX Implementation Details}
\label{app:jax_implementation}
 
The results presented in Section~\ref{sec:results} were obtained using a NumPy-PyTorch implementation in which the geometric quantities (strain Jacobians $\mathbf{G}_k$ and Hessians $\mathbf{H}_{\epsilon,k}$) are computed analytically in NumPy while the constitutive derivatives ($\partial\mathcal{E}_k/\partial\boldsymbol{\epsilon}_k$, $\nabla^2_{\boldsymbol{\epsilon}_k}\mathcal{E}_k$) are obtained via PyTorch autograd. This hybrid approach, while modular and convenient for prototyping diverse energy models, incurs overhead from (i) repeated data transfers between NumPy and PyTorch tensor formats at each Newton iteration, (ii) a Python-level loop over the four strain components when computing the constitutive Hessian via \texttt{torch.autograd.grad}, (iii) the inability to fuse the geometric and constitutive stages of the chain-rule assembly into a single compiled kernel, since they execute in separate frameworks, and (iv) dense $\mathcal{O}(N_\mathrm{DOF}^{2})$ Hessian assembly and factorization in the linear solver, despite the underlying sparsity pattern being banded with fixed bandwidth.
 
To quantify the achievable performance of the Discrete Elastic Ribbon framework and to provide a fair efficiency comparison across energy models, we developed a complete reimplementation in JAX~\cite{jax2018github} with Equinox~\cite{kidger2021equinox}. Equinox provides a PyTorch-like module abstraction within the JAX ecosystem, enabling us to represent energy models, rod geometry, and boundary conditions as composable pytree objects that remain fully compatible with JAX's functional transformations (\texttt{jit}, \texttt{vmap}, \texttt{grad}). The JAX implementation preserves the same algorithmic structure at the Newton-Raphson level (implicit Euler integration, Newton-Raphson iteration with adaptive time-stepping, and adaptive Tikhonov regularization with singular-value fallback per Algorithm~\ref{alg:robust_solve}) but replaces the reference's dense Hessian assembly and dense LAPACK factorization with direct assembly into LAPACK banded storage (bandwidth $k = 10$, inherited from the 11-DOF triplet stencil) and a banded LU factorization via \texttt{scipy.linalg.solve\_banded}. Both assembly and the Newton-step linear solve therefore scale as $\mathcal{O}(N)$ in the number of centerline nodes, matching the $\mathcal{O}(N)$ cost predicted by the banded structure of the discrete elastic ribbon Hessian. The reimplementation exploits four key capabilities:
 
\begin{enumerate}
    \item \textbf{Vectorized assembly via \texttt{vmap}}: The per-element energy, gradient, and Hessian computations are batched across all $N-2$ triplet stencils using \texttt{jax.vmap}, replacing the NumPy-PyTorch batch operations with a single compiled vectorized kernel that fuses the constitutive and geometric derivative stages. Unlike the reference implementation, which requires separate batched calls to PyTorch (for constitutive derivatives) and NumPy (for the chain-rule assembly), the \texttt{vmap}'d kernel executes the entire per-element computation (strain evaluation, energy differentiation, and chain-rule assembly) in a single pass.
    \item \textbf{JIT compilation}: The entire Newton-Raphson iteration is compiled into a single \texttt{jax.lax.while\_loop} body, eliminating Python interpreter overhead from the inner loop. The only host round-trip per Newton iteration is the banded factorization via \texttt{jax.pure\_callback} (see item~4 below). All other operations (assembly, residual evaluation, convergence checks, regularization escalation) remain within a single XLA-compiled kernel.
    \item \textbf{Hybrid analytical-autodiff derivatives}: Both implementations use the same hybrid derivative strategy: analytical strain Jacobians $\mathbf{G}_k$ and Hessians $\mathbf{H}_{\epsilon,k}$, with autodiff for the constitutive derivatives (PyTorch \texttt{autograd} in the reference, \texttt{jax.grad} and \texttt{jax.hessian} in JAX). The key advantage of the JAX version is that all operations, geometric and constitutive, reside within a single tracing framework, allowing the XLA compiler to fuse the chain-rule assembly (Steps~1--3 of Algorithm~\ref{alg:general_energy}) into one optimized kernel without intermediate array materialization or cross-framework data transfers.
    \item \textbf{Banded linear algebra}: The per-element $11 \times 11$ Hessian blocks produced by the \texttt{vmap}'d assembly kernel are scattered directly into LAPACK banded storage of shape $(2k+1, N_\mathrm{DOF})$ rather than into an $N_\mathrm{DOF}\times N_\mathrm{DOF}$ dense matrix, eliminating the $\mathcal{O}(N_\mathrm{DOF}^{2})$ zero-initialization and scatter-into-dense cost that otherwise dominates at modest $N$. Dirichlet boundary conditions are enforced by adding a large diagonal penalty at fixed DOFs, which preserves the banded sparsity pattern of the stiffness matrix end-to-end. This avoids the bandwidth-breaking principal-submatrix extraction used by the dense path. The banded factorization itself is dispatched to the host via \texttt{jax.pure\_callback}, since \texttt{jax.scipy.linalg} does not currently expose a banded solver; the callback overhead ($\sim$140\,\textmu s per call at our problem sizes) is dominated by the linear-solve cost savings even at $N = 45$.
\end{enumerate}
 
The energy models, rod geometry, and boundary conditions are implemented as \texttt{equinox.Module} pytrees, making the per-step Newton solve differentiable via the implicit function theorem applied to the equilibrium conditions. A fully JIT-compiled simulation mode (\texttt{jax.lax.scan} over time steps) is also available for fixed boundary-condition schedules, enabling gradient-based inverse design and parameter estimation. While these differentiable capabilities are not exploited in the present benchmarks and energy models, they demonstrate the extensibility of the framework toward data-driven applications discussed in Section~\ref{sec:conclusion}.

\bibliographystyle{unsrt}   
\bibliography{references} 

\end{document}